\documentclass[apjl]{emulateapj}
\usepackage{psfig,amsfonts,amsmath,graphicx,natbib,apjfonts,subfigure}

\def\ra#1#2#3{#1$^{\rm h}$#2$^{\rm m}$#3$^{\rm s}$}
\def\dec#1#2#3{#1$^\circ$#2$'$#3$''$}
\def\nod{\nodata}
\def\swift{{\it Swift}}
\def\chandra{{\it Chandra}}
\def\grb{GRB\,111020A}

\def\har{1}
\def\nasa{2}
\def\psu{3}
\def\ox{4}

\begin{document}

\title{A Jet Break in the X-ray Light Curve of Short \grb:
Implications for Energetics and Rates}

\author{
W.~Fong\altaffilmark{\har}, E.~Berger\altaffilmark{\har},
R.~Margutti\altaffilmark{\har}, B.~A.~Zauderer\altaffilmark{\har},
E.~Troja\altaffilmark{\nasa}, I.~Czekala\altaffilmark{\har},
R.~Chornock\altaffilmark{\har}, N.~Gehrels\altaffilmark{\nasa},
T.~Sakamoto\altaffilmark{\nasa}, D.~B.~Fox\altaffilmark{\psu},
P.~Podsiadlowski\altaffilmark{\ox}
}

\altaffiltext{\har}{Harvard-Smithsonian Center for Astrophysics, 60
Garden Street, Cambridge, MA 02138, USA}

\altaffiltext{\nasa}{NASA Goddard Space Flight Center, Greenbelt, MD 20771, USA}

\altaffiltext{\psu}{Department of Astronomy and Astrophysics, 525
Davey Laboratory, Pennsylvania State University, University Park, PA
16802, USA}

\altaffiltext{\ox}{Department of Astronomy, Oxford University, Oxford, OX1 3RH, UK}

\begin{abstract}

We present broad-band observations of the afterglow and environment of
the short GRB\,111020A. An extensive X-ray light curve from
\swift/XRT, XMM-Newton and \chandra, spanning $\sim 100$ seconds to
$10$ days after the burst, reveals a significant break at $\delta t
\approx\!2$ days with pre- and post-break decline rates of
$\alpha_{X,1}\approx -0.78$ and $\alpha_{X,2} \lesssim -1.7$,
respectively. Interpreted as a jet break, we infer a collimated
outflow with an opening angle of $\theta_j \approx 3-8^{\circ}$. The
resulting beaming-corrected $\gamma$-ray ($10-1000$ keV band) and
blastwave kinetic energies are $(2-3) \times 10^{48}$ erg and $(0.3-2)
\times 10^{49}$ erg, respectively, with the range depending on the
unknown redshift of the burst. We report a radio afterglow limit of
$<\!39$ $\mu$Jy ($3\sigma$) from EVLA observations which, along with
our finding that $\nu_c<\nu_X$, constrains the circumburst density to
$n_0 \sim 0.01-0.1$ cm$^{-3}$. Optical observations provide an
afterglow limit of $i \gtrsim 24.4$ mag at $18$ hours after the burst,
and reveal a potential host galaxy with $i \approx 24.3$ mag. The
sub-arcsecond localization from \chandra\ provides a precise offset of
$0.80'' \pm 0.11''$ ($1\sigma$) from this galaxy corresponding to an
offset of $5-7$ kpc for $z=0.5-1.5$. We find a
high excess neutral Hydrogen column density of $(7.5 \pm 2.0) \times
10^{21}$ cm$^{-2}$ ($z=0$). Our observations demonstrate that a
growing fraction of short GRBs are collimated which may lead to a true
event rate of $\gtrsim 100-1000$ Gpc$^{-3}$ yr$^{-1}$, in good
agreement with the NS-NS merger rate of $\approx 200-3000$
Gpc$^{-3}$ yr$^{-1}$. This consistency is promising for coincident
short GRB-gravitational wave searches in the forthcoming era of
Advanced LIGO/VIRGO.

\end{abstract}

\section{Introduction}

Observations of the temporal and spectral evolution
of short-duration gamma-ray burst (GRB; $T_{90}<2$ sec;
\citealt{kmf+93}) afterglows are crucial to our understanding of the basic
properties of these events: their energetics, parsec-scale
environments, and geometries. From observations over the past $7$
years, we now know that short GRBs have isotropic-equivalent energies
of $\sim 10^{50}-10^{52}$ erg \citep{ber07} and circumburst densities
of $\sim 10^{-6}-1$ cm$^{-3}$
\citep{sbk+06,pan06,sdp+07,pmg+09,ber10,fbc+11}; however these ranges
are based on only a handful of events. The geometry, or degree of
collimation, is the least constrained property, but is of particular
interest because it directly affects the true energy scale and event
rates. These parameters aid our understanding of
the explosion physics, the nature of the progenitors, and the
potential detectability of short GRBs as gravitational wave
sources. In particular, knowledge of the true energy scale may
constrain the mechanism of energy extraction from the central engine
and the ejecta composition: $\nu\bar{\nu}$ annihilation powering a
baryonic jet \citep{jar93,mhi+93} or magnetohydrodynamic (MHD)
processes in a magnetically-dominated outflow
\citep{bz77,rrd03}. Significant improvement on the short GRB observed
rate of $\gtrsim 10$ Gpc$^{-3}$ yr$^{-1}$ \citep{ng07} will have a
critical impact on estimates for coincident short GRB-gravitational
wave detections in the era of Advanced LIGO/VIRGO \citep{aaa+10}.

The opening angles ($\theta_j$) of GRBs can be inferred from temporal
breaks in the afterglow light curves (``jet breaks''), which occur at
the time, $t_j$, when the Lorentz factor of the outflow is
$\Gamma(t_j)\!\approx\!1/\theta_j$; a later break corresponds to a
wider opening angle \citep{sph99,rho99}. Jet breaks in the light
curves of long-duration GRBs have led to an opening angle distribution
with a range of $\sim\!2-20^\circ$ and a median of $7^{\circ}$,
leading to beaming-corrected energies of $E_{\gamma}=[1-{\rm
cos}(\theta_j)]E_{\gamma,{\rm iso}} \sim 10^{50}-10^{51}$ erg
\citep{bfk03,fks+01,fb05,kb08,rlb+09}. For short GRBs, there is
mounting theoretical \citep{elp+89,npp92} and observational
\citep{fbf10,ber10,cld+11} evidence that the progenitors are
NS-NS/NS-BH mergers and numerous simulations of post-merger black hole
accretion have predicted collimated outflows with $\theta_j \sim
5-20^{\circ}$ \citep{pwf99,ajm+05,ros05,rgb+11} up to several tens of
degrees \citep{rj99a,pwf99,rgb+11}.

However, the detection of jet breaks in the afterglow light curves of
short GRBs has proved to be challenging. They can in principle be
measured from optical or radio observations, but there are several
caveats that make this particularly difficult for short GRBs.  First,
the brightness of optical and radio afterglows are sensitive to the
circumburst densities, which are typically low, $\sim 10^{-2}$
cm$^{-3}$ \citep{sbk+06}. Indeed, of nearly $70$ short bursts detected
by \swift, only 2 radio afterglows have been detected over the past
$7$ years \citep{bpc+05,sbk+06,cf11}. Similarly, only $\sim 30\%$ of
\swift\ bursts have detected optical afterglows, with a typical
brightness at $\lesssim 1$ day of $\approx\!23$ mag
\citep{ber10,fbc+11}, making long-term temporal monitoring nearly
impossible with ground-based facilities. Second, in the optical band
there can be significant contamination from the host galaxies, which
are generally brighter than the afterglows at $\gtrsim\!1$ day
\citep{ber10}.

On the other hand, the X-ray afterglow brightness is independent of
the circumburst density (as long as the density is $\gtrsim\!10^{-5}$
cm$^{-3}$ and hence $\nu_c>\nu_X$; \citealt{gs02}), and host
contamination is not an issue. In addition, the well-sampled
\swift/XRT light curves from $\sim\!1$ min to $\sim\!1$ day provide an
unambiguous baseline against which we can measure a subsequent
break. Therefore, it is no surprise that the X-rays enabled the
discovery of the first jet break in a short GRB. The X-ray afterglow
light curve of GRB\,051221A exhibited a break at $\approx\!5$ days,
leading to $\theta_j\approx\!7^{\circ}$
\citep{sbk+06,bgc+06}. Similarly, \chandra\ observations of
GRB\,050724A out to $22$ days placed a meaningful lower limit of
$\theta_j\!\gtrsim\!25^\circ$ \citep{gbp+06}, consistent with a
spherical explosion. Temporal breaks on timescales of $\lesssim$ few
hours were observed in the XRT light curves of GRBs\,061201
\citep{sdp+07} and 090510\footnotemark\footnotetext{GRB\,090510 also
exhibits a post-jet break-like decay in its optical light curve
\citep{nkk+12}.} \citep{dsk+10}. If these are interpreted as jet
breaks, they would lead to $\theta_j \sim 1^{\circ}$; however, they
also match the timescale and behavior of early breaks in long GRBs,
which are not due to collimation \citep{nkg+06,zfd+06,lzz07}. Finally, there
is tentative evidence for beaming in the light curves of GRBs\,050709
\citep{ffp+05} and 061210\footnotemark\footnotetext{Please note that
\citet{ber07} erroneously refers to GRB\,061006.} \citep{ber07};
however, these are based on sparsely-sampled light curves without a
definitive break (e.g., \citealt{whj+06}).  No other jet breaks in the
light curves of unambiguous short GRBs have been reported to
date\footnotemark\footnotetext{A jet break was reported in the light
curve of GRB\,090426A \citep{nkr+11}; however the characteristics of
its environment and prompt emission are more similar to those of long
GRBs \citep{lbb+10,xlw+11}.} and the lack of jet breaks in \swift/XRT
observations out to $\sim\!1-2$ days can provide only weak lower
bounds of $\theta_j\!\gtrsim\!2-6^\circ$ (revised from
\citealt{chp+12} with more realistic density values; see
Section~\ref{sec:ber}).

Against this backdrop, we present here the discovery of a break in the
X-ray light curve of \grb at $\delta t \approx 2$ days, best explained
as a jet break. We also present contemporaneous radio and optical
limits on the afterglow, allowing a characterization of the broad-band
synchrotron spectrum and constraints on the energy and density. In
addition, we report the discovery of a putative host galaxy. A
comparison of our X-ray and optical data may require an appreciable
amount of extinction and the highest intrinsic neutral Hydrogen column
density for a short GRB to date. Our results have strong implications
for the opening angle distribution, and therefore the observed short
GRB rate and true energy release.

Unless otherwise noted, all magnitudes in this paper are in the AB system
and are corrected for Galactic extinction in the direction of the
burst using $E(B-V)=0.432$ mag \citep{sfd98,sf11}. We
employ a standard $\Lambda$CDM cosmology with $\Omega_M=0.27$,
$\Omega_\Lambda=0.73$, and $H_0=71$ km s$^{-1}$ Mpc$^{-1}$.

\section{Observations of \grb}

\subsection{\swift\ Observations}

\grb\ was detected on 2011 October 20 at 06:33:49.0 UT by the Burst
Alert Telescope (BAT) on-board the \swift\ satellite
\citep{ggg+04,gcn12460}. BAT located the burst at a ground-calculated
position of RA=\ra{19}{08}{06.9} and Dec=$-$\dec{38}{01}{50.3} (J2000)
with $2.1'$ accuracy ($90\%$ containment; \citealt{gcn12464}). The
X-ray Telescope (XRT) commenced observations of the location of the
burst at $\delta t=72.8$ s (where $\delta t$ is the time after the BAT
trigger) and detected a fading X-ray source
(Section~\ref{sec:XRTobs}). The UV-Optical Telescope (UVOT) began
observations of the field at $\delta t=79$ s but no corresponding UV
or optical source was found within the XRT position. The $3\sigma$
limit in the $white$ filter, which transmits over
$\lambda=1600$-$8000$ \AA\ \citep{pbp+08}, is $20.3$\,mag (not
corrected for Galactic extinction; \citealt{gcn12466}).

The gamma-ray emission consists of a single pulse with a duration of
$T_{90}=0.40 \pm 0.09$ s in the $15-350$ keV band, classifying \grb\
as a short burst \citep{gcn12464}. The spectrum is best fit with a
single power law with index $1.37 \pm 0.26$ and a fluence of
$f_\gamma=(6.5 \pm 1.0) \times 10^{-8}$ erg cm$^{-2}$ ($15-150$
keV). Spectral lag analysis is not conclusive, and there is no clear
evidence for extended emission \citep{gcn12477}.

\subsection{X-ray Observations}
\label{sec:XRTobs}

The XRT promptly located a fading, uncatalogued X-ray
source \citep{ebp+07,ebp+09, gcn12460} with a UVOT-enhanced position of
RA=\ra{19}{08}{12.53} and Dec=$-$\dec{38}{00}{43.8} (J2000) and an
uncertainty of $1.6''$ \citep{gcn12463}. XRT observations of the field
continued until the source faded below the detection threshold at
$\delta t \approx 3.5$ days.

We also observed the field of \grb\ with the European Photon Imaging
Camera (EPIC-PN) on-board the X-ray Multi-Mirror Mission (XMM-Newton)
starting at $\delta t=0.65$ days. With $13.5$ ks of on-source
observations, we clearly detect a source in the energy range $0.5-10$
keV, consistent with the \swift-XRT position. In addition, we obtained
two sets of $20$-ks observations with the Advanced CCD Imaging
Spectrometer (ACIS-S; $0.3-10$\,keV) on-board the {\it Chandra} X-ray
Observatory at $\delta t=2.9$ and $10.1$ days to refine the astrometry
and monitor the light curve evolution. We detect the X-ray afterglow
in the first \chandra\ observation but do not detect any source at the
same location in the second epoch.

\subsubsection{Data Analysis and Spectral Fitting}
\label{sec:specfit}

We analyze the XRT data using the latest version of the HEASOFT
package (v.6.11) and relevant calibration files. We apply standard
filtering and screening criteria, and generate a count rate light
curve following the prescriptions from \citet{mgg+10} and
\citet{mzb+12}. Our re-binning scheme ensures a minimum
signal-to-noise ratio of $S/N=4$ for each temporal bin. We analyze the
XMM data using standard routines in the Scientific Analysis System
(SAS) v.11. We remove the first $5$ ks of data due to high
background contamination, giving a total exposure time of $13.5$
ks. We extract count rates from a $20''$ radius aperture and the
background is calculated using $20''$ radius source-free regions on
the same chip. We use the {\tt CIAO} data reduction package for the {\it
Chandra} data. For the first epoch, we use a $2.5''$ radius source
aperture centered on the {\it Chandra} position and a background
annulus with inner and outer radii of $10''$ and $35''$, respectively,
giving a source detection significance of $\sim 30\sigma$. For the
second epoch, we extract $1$ count in a $2.5''$ aperture at the
location of the source, consistent with the average background level
calculated from source-free regions on the same chip. We therefore
take the $3\sigma$ background count rate as an upper limit on the
X-ray afterglow.

To extract a spectrum from the X-ray data, we fit each of the data
sets with an absorbed power law model ($tbabs \times ztbabs \times
pow$ within the XSPEC routine) characterized by a photon index,
$\Gamma$, and intrinsic neutral hydrogen absorption column,
$N_{\rm{H,int}}$, in excess of the Galactic column density in the
direction of the burst, $N_{{\rm H,MW}}=6.9\times
10^{20}\,\rm{cm^{-2}}$ (typical uncertainty of $\sim 10\%$;
\citealt{kbh+05,wlb11}), using Cash statistics. For XRT, we utilize
data in the time interval $0.08-60$ ks where there is no evidence for
spectral evolution. We find an average best-fitting (C-stat$_\nu=0.86$
for $188$ d.o.f.) spectrum characterized by $\Gamma=2.2 \pm 0.5$ and
$N_{\rm{H,int}}=(1.0 \pm 0.3) \times 10^{22}\,\rm{cm^{-2}}$ at $z=0$
(Table~\ref{tab:xrayspec}). Uncertainties correspond to the $90\%$
confidence level. Our best-fit parameters are fully consistent with
the automatic spectrum fit produced by \citet{gcn12468}. The XMM data
are best modeled with a power law characterized by $\Gamma=2.0 \pm
0.4$ and $N_{\rm H,int} = (0.65 \pm 0.22) \times 10^{22}$ cm$^{-2}$
(C-stat$_\nu=1.0$ for $256$ d.o.f.), consistent with the XRT model
parameters. We also fit the first epoch of {\it Chandra} data and the
resulting parameters are consistent with those from XRT and XMM,
albeit with large error bars due to low count statistics
(Table~\ref{tab:xrayspec}). Since we find no evidence for spectral
evolution in the XRT data, we perform a joint XRT+XMM spectral
analysis to obtain the best constraints on $\Gamma$ and $N_{\rm
H,int}$. The resulting best-fit model has $\Gamma=2.0 \pm 0.3$ and
$N_{\rm{H,int}}=(0.8 \pm 0.2)\times 10^{22}$ cm$^{-2}$ ($90\%$ c.l.;
C-stat$_\nu=0.94$ for 446 d.o.f.). Although the redshift of the burst
is unknown, we note that $\Gamma$ remains unchanged within its
$1\sigma$ value for $z \lesssim 3$ and we find evidence for intrinsic
$N_{\rm H,int}$ in excess of the Galactic value at $6.5\sigma$
confidence. The best-fitting spectral parameters for each of the three
data sets and the joint fit are summarized in
Table~\ref{tab:xrayspec}.

Applying these parameters to the individual XRT, XMM, and {\it
Chandra} data sets, we calculate the count rate-to-flux conversion
factors, and hence their absorbed and unabsorbed fluxes
(Table~\ref{tab:xrayobs}). Applying these conversion factors results
in the X-ray light curve shown in Figure~\ref{fig:xraylc}.

\subsubsection{Differential Astrometry}
\label{sec:astrometry}

In the absence of the detection of an optical afterglow
(Section~\ref{sec:GMOS}), we use our \chandra\ observations to refine
the \swift/XRT position ($1.6''$ uncertainty) to sub-arcsecond
accuracy. We perform differential astrometry between our {\it Chandra}
and GMOS observations (Section~\ref{sec:GMOS}). To achieve the
maximum signal-to-noise ratio, we combine both epochs of GMOS $i$-band
observations and use {\tt
SExtractor}\footnotemark\footnotetext{http://sextractor.sourceforge.net/.}
to determine the positions and centroid uncertainty of sources in the
field. Performing an absolute astrometric tie to 2MASS using $\sim$70
common point sources, we find a resulting rms value of $\sigma_{\rm
GMOS-2MASS}$ = $0.17''$ ($0.12''$ in each coordinate).

To refine the native \chandra\ astrometry and determine the location
of the X-ray afterglow relative to the GMOS image, we perform
differential astrometry. We use {\tt CIAO} routines {\tt mergeall} to
combine the two {\it Chandra} epochs and {\tt wavdetect} to obtain
positions and $1\sigma$ centroid uncertainties of X-ray sources in the
field. We also use {\tt wavdetect} to determine the {\it Chandra}
position of the afterglow from the first epoch and find a $1\sigma$
centroid statistical uncertainty $\sigma_{{\rm X,ag}} = 0.08''$. We
calculate an astrometric tie based on two X-ray and optically bright
common sources and find weighted mean offsets of $\delta$RA$=-0.27''
\pm 0.06''$ and $\delta$Dec$=+0.05'' \pm 0.05''$ giving a tie
uncertainty of $\sigma_{\rm CXO-GMOS}=0.08''$. There are three
additional common, but fainter sources. An astrometric tie using all
five sources gives weighted mean offsets and errors of
$\delta$RA$=-0.29'' \pm 0.15''$ and $\delta$Dec$=+0.06'' \pm 0.16''$,
fully consistent with our results from using the two bright sources
alone. We therefore use the astrometric solution from the two bright
sources only. Applying this solution, we obtain a {\it Chandra} X-ray
afterglow position of RA=\ra{19}{08}{12.49} and
Dec=$-$\dec{38}{00}{42.9} (denoted by the blue circle in
Figure~\ref{fig:gmos_psf_panel}) with a total $1\sigma$ uncertainty of
$0.20''$, accounting for the 2MASS-GMOS astrometric tie, GMOS-{\it
Chandra} tie, and afterglow positional uncertainty. This position is
consistent with the XRT position and is offset by $0.94''$ from the
XRT centroid.

\subsection{Optical Observations and Putative Host Galaxies}
\label{sec:GMOS}

\begin{figure}
\centering
\includegraphics[angle=0,width=6.4in]{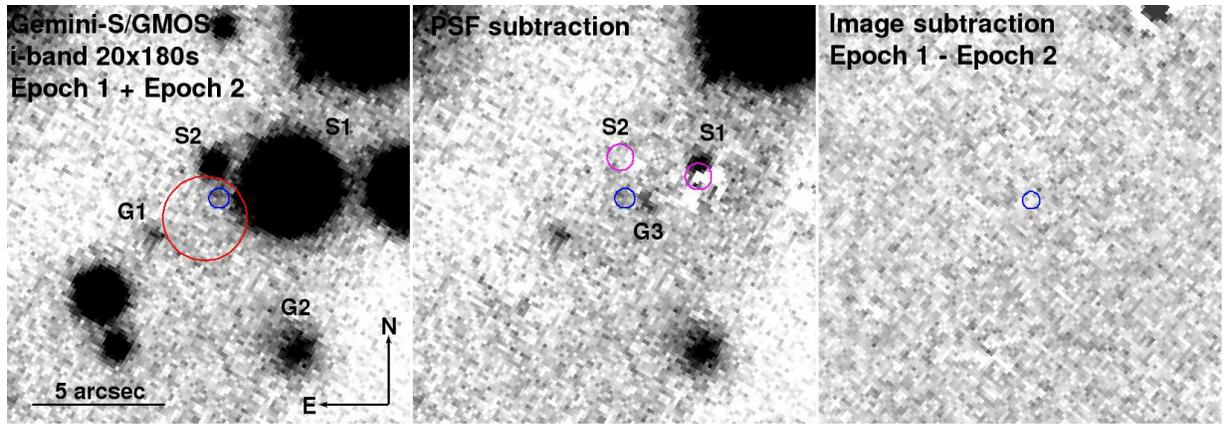}
\caption{Optical $i$-band observations obtained with GMOS on
Gemini-South. {\it Left}: Combined stack of two nights of GMOS
$i$-band data. Stars S1 and S2 are labeled, as well as galaxies G1 and
G2. X-ray positions of \grb\ are denoted by the circles (red: {\it
Swift}-XRT, $1.6''$ radius, $90\%$ containment; blue: {\it Chandra},
$0.33''$ radius, $90\%$ confidence). {\it Center}: PSF-subtracted
image with the centroids of S1 and S2 (magenta circles). The
subtraction reveals a third source, G3, with $i \approx 24.3$
mag. {\it Right}: Digital image subtraction of the two epochs obtained
at $17.7$ hours and $1.7$ days, respectively, reveals no residuals in
or around the {\it Chandra} position.
\label{fig:gmos_psf_panel}} 
\end{figure}

We initiated $i$-band observations of \grb\ with the Gemini
Multi-Object Spectrograph (GMOS) mounted on the Gemini-South 8-m
telescope on 2011 October 21.01 UT ($\delta t=17.7$ hours). The data
were reduced using the {\tt gemini/gmos} package in IRAF. In a stack
of $9 \times 180$ s exposures in $0.74''$ seeing and photometric conditions
(Figure~\ref{fig:gmos_psf_panel}), we do not detect any sources within
the enhanced XRT error circle or coincident with the {\it Chandra}
position. However, the outskirts of the XRT position are partially
contaminated by emission from a nearby $i=17.7$ mag star (S1) and a
fainter star (S2) with $i = 22.7$ mag (corrected for $A_i=0.73$ mag;
Figure~\ref{fig:gmos_psf_panel}). We detect two additional nearby
sources: a faint galaxy (G1) located $2.8''$ away from the center of
the \chandra\ position and a brighter galaxy (G2) with a $6.5''$
offset (Figure~\ref{fig:gmos_psf_panel}).

To search for a fading optical afterglow, we obtained a second, deeper
set of $i$-band observations ($11 \times 180$ s) with GMOS at $\delta
t=1.74$ days in $0.67''$ seeing.  Digital image subtraction using the
ISIS software package \citep{ala00} reveals no variation between the
two epochs inside or near the X-ray afterglow error circles
(Figure~\ref{fig:gmos_psf_panel}). To calculate the upper limit on the
afterglow, we add several point sources of varying magnitudes between
$i=24-26$ mag around the position in the first epoch using IRAF
routine {\tt addstar}. We perform photometry in the residual image in
$2''$ apertures using the standard published $i$-band zeropoint for
GMOS-S and place a $3\sigma$ limit on the afterglow of $i \gtrsim 24.4$ mag
($F_\nu \lesssim 0.63$ $\mu$Jy). We also perform photometry in a $1.8''$
aperture for G1 and a $2.3''$ aperture for G2, giving magnitudes of
$i=23.9 \pm 0.2$ mag and $i=21.9 \pm 0.1$, respectively
(Table~\ref{tab:opt_phot}).

In addition, we obtained $r$-band observations ($3 \times 360$ s in
$0.62''$ seeing) with the Low Resolution Survey Spectrograph~3
(LDSS3) mounted on the Magellan/Clay $6.5$-m telescope concurrent to
the first epoch of GMOS observations ($\delta t=17.7$ hours). We
easily detect G2, with $r=21.1 \pm 0.1$ mag, but do not detect G1 to a
$3\sigma$ limit of $r \gtrsim 23.4$ (corrected for $A_r=0.99$ mag;
Table~\ref{tab:opt_phot}), where the zeropoint is determined from
several standard stars at similar airmass.

Since S1 and S2 contaminate the {\it Chandra} position, we subtract
their contribution using point-spread-function (PSF) subtraction on
the individual observations and a combined stack of the two GMOS
epochs. We use standard PSF-fitting routines in the IRAF {\tt daophot}
package. Modelling the PSF using 4 bright, unsaturated stars in the
field out to a radius of $3''$ ($\sim 4\theta_{\rm FWHM}$) from the
center of each star, we subtract several stars in the field including
S1 and S2. The clean subtraction of these stars indicates a model PSF
representative of the PSF of the field. We uncover a faint, mildly
extended source (G3) on the outskirts of S1 at coordinates
RA=\ra{19}{08}{12.43} and Dec=$-$\dec{38}{00}{43.07} (J2000). This
source, which lies $0.80''$ from the center of the {\it Chandra} error
circle, has a magnitude of $i=24.3 \pm 0.2$ and is a potential host of
\grb\ (Section~\ref{sec:probcc}). However, we cannot rule out the
possibility that this source is a faint star.

\subsection{Radio Observations and Possible Afterglow}
\label{sec:evla}

We observed the position of \grb\ with the Expanded Very Large
Array\footnotemark\footnotetext{Newly renamed the Karl~G.~Jansky Very
Large Array.}  (EVLA) beginning on 1 October 20.95 UT ($\delta t=16.1$
hours; Program $10C-145$) at a mean frequency of $5.8$ GHz with a
total on-source integration time of $65$ min. We observed 3C295 and
J1937$-$1958 for bandpass/flux and gain calibration, respectively, and
used standard procedures in the Astronomical Image Processing System
(AIPS; \citealt{gre03}) for data calibration and analysis. With the
new wideband capabilities of the EVLA \citep{pcb+11}, our data have an
effective bandwidth of $\sim$1.5 GHz after excising edge channels and
data affected by radio frequency interference. The low declination of
\grb\ and the compact D configuration of the array at the time of
observation caused significant shadowing and required the removal of 7
out of 27 antennas (the north arm of the EVLA).

Taking into account the highly-elongated beam ($33'' \times 7''$ with
a position angle of $170^\circ$), we detect a $3.7\sigma$ source with
an integrated flux density of $48 \pm 13\,\mu$Jy located at
RA=\ra{19}{08}{12.40}, Dec=$-$\dec{38}{00}{41.2} ($\delta$RA$=1.1''$,
$\delta$Dec$=3.6''$, $1\sigma$ uncertainty), consistent with the {\it
Chandra} position. The position, peak flux and integrated flux of the
source are consistent regardless of our choice of weighting, or if we
analyze the upper and lower sidebands separately. However, despite the
statistical significance of the detection, we cannot completely rule
out sidelobe contribution from nearby bright sources in the field due
to the low declination of the burst. Therefore, we conservatively
adopt a $3\sigma$ upper limit of $39\,\mu$Jy on the radio afterglow of
\grb\ for our analysis. We note that if the source is indeed real then
upper limits inferred from the radio data can be treated as actual
values. \\ \\

\section{Results}

\subsection{Galaxy Probabilities of Chance Coincidence}
\label{sec:probcc}

To assess which galaxy is the most probable host of \grb\, we adopt
the methodology of \citet{bkd02} and \citet{ber10} to calculate the
probability of chance coincidence $P(<\delta R)$ at a given angular
separation $\delta R$. We determine the expected number density of
galaxies brighter than a measured apparent magnitude, $m$, using the
results of deep optical galaxy surveys \citep{hpm+97,bsk+06}:

\begin{equation}
\sigma(\le m)=\frac{1}{0.33\times {\rm ln}(10)}\times
10^{0.33(m-24)-2.44} \,\,\,\,{\rm arcsec}^{-2}.
\label{eqn:gal}
\end{equation}

\noindent Then the probability of chance coincidence is given by \citep{bkd02}

\begin{equation}
P(<\delta R)=1-{\rm e}^{-\pi (\delta R)^2\sigma(\le m)}.
\label{eqn:prob}
\end{equation}

\begin{figure}
\centering
\includegraphics[angle=0,width=6.4in]{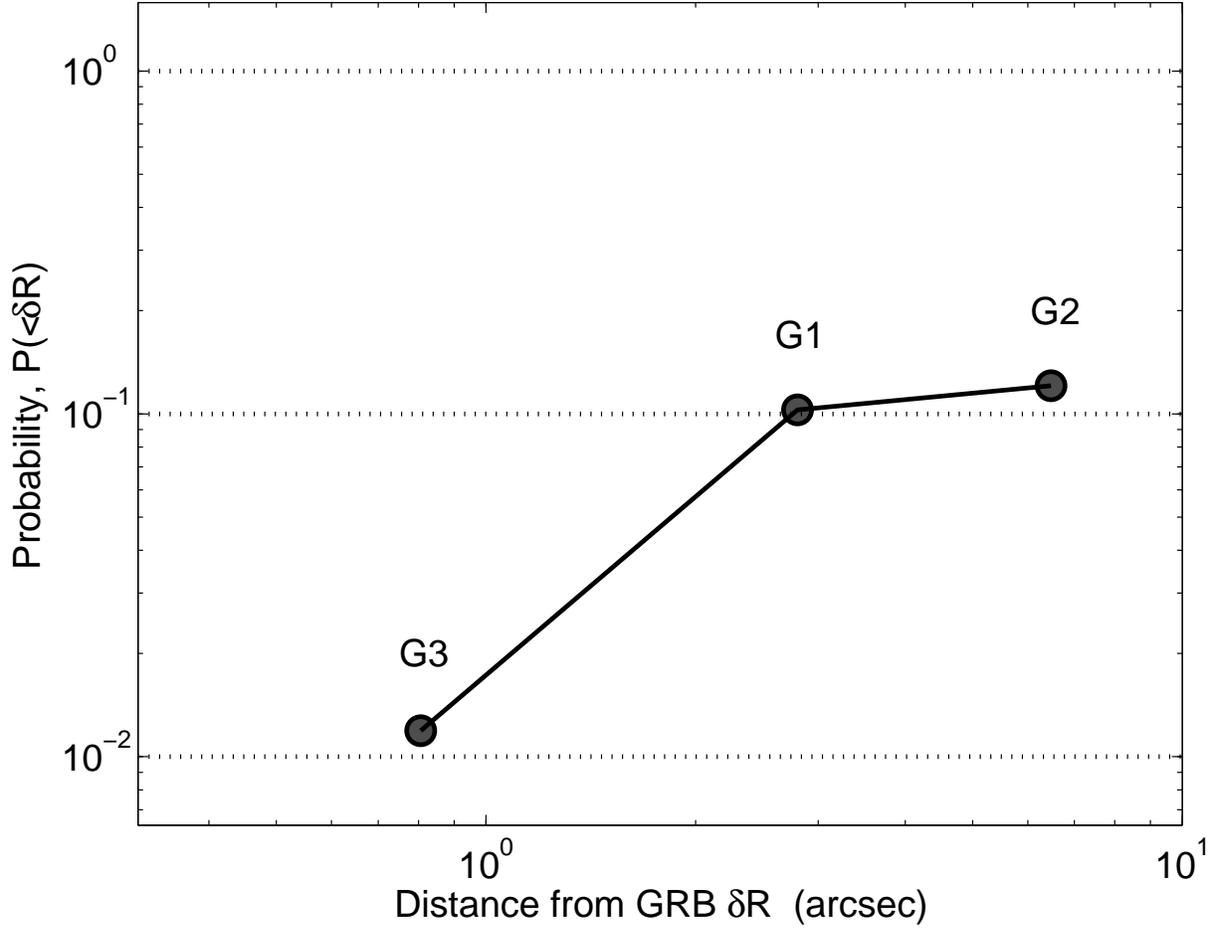}
\caption{Probability of chance coincidence, $P(<\delta R)$, as a
function of angular distance from the center of the {\it Chandra}
afterglow position for the three host galaxy candidates of \grb. The
galaxy G3 has the lowest probability of chance coincidence $P(<\delta
R)=0.01$, and is therefore the most probable host of \grb.
\label{fig:111020a_cc}} 
\end{figure}

We calculate $P(<\delta R)$ for each of the three host galaxy
candidates (Figure~\ref{fig:111020a_cc}), and find that G3 is the most
probable host of \grb\ with $P(<\delta R)=0.01$, while for G1 and G2,
the values are $P(<\delta R)=0.10$ and $0.12$, respectively. \\ \\ \\

\subsection{X-ray Light Curve Fitting and a Jet Break}
\label{sec:jb}

\begin{figure}
\centering
\includegraphics[angle=0,width=6.4in]{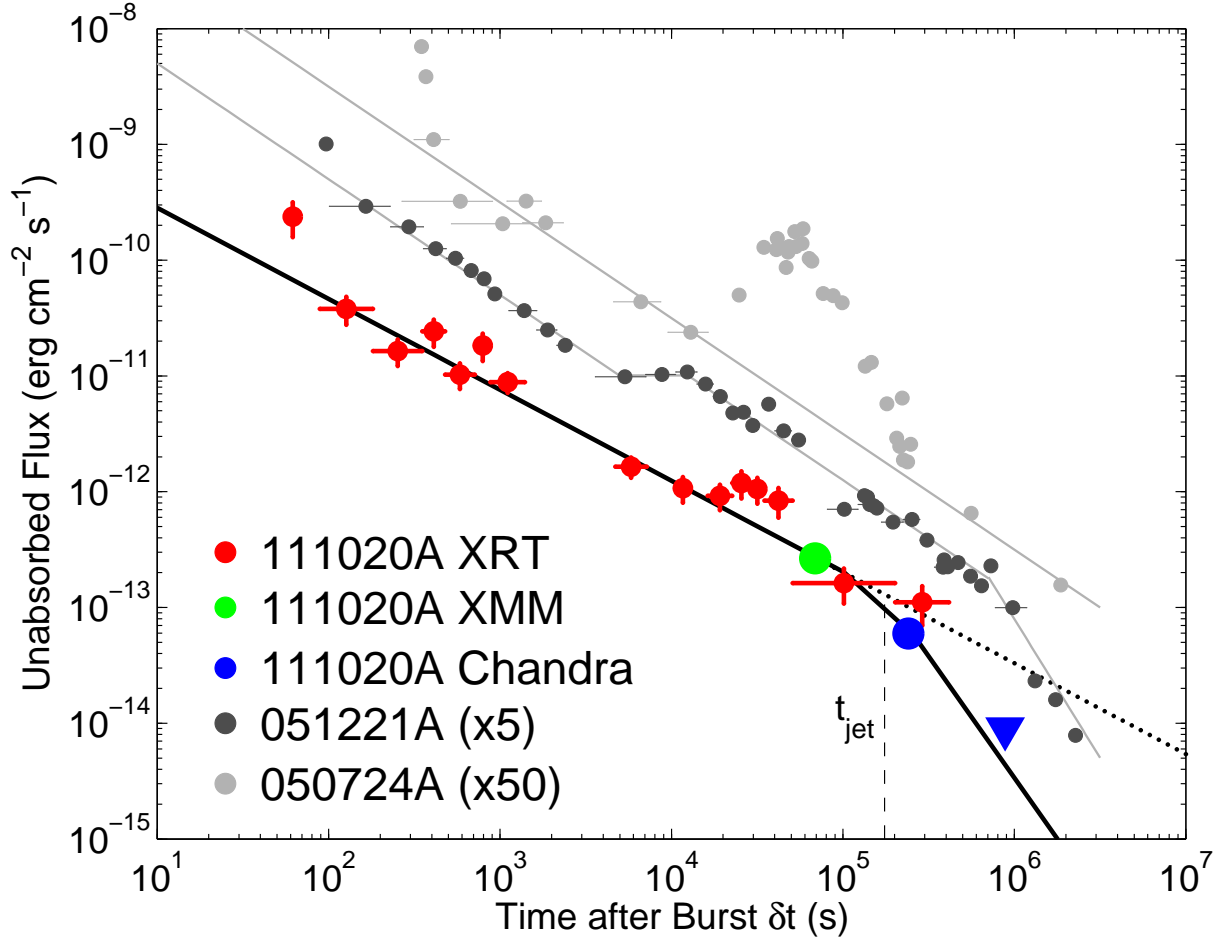}
\caption{Unabsorbed X-ray flux light curve for \grb\ from \swift-XRT
(red), XMM (green), and {\it Chandra} (blue). Flux errors are
$1\sigma$. The {\it Chandra} $3\sigma$ upper limit is denoted by the
blue triangle. The best-fit broken power law model (black solid line)
for \grb\ is characterized by $\alpha_1=-0.78$, $\alpha_2=-2.1$, and
$t_j=2.0$ days. A single power law model with $\alpha=-0.78$ (black
dotted) violates the {\it Chandra} upper limit. Also plotted are X-ray
light curves for short GRBs\,051221A (dark grey circles;
\citealt{sbk+06,bgc+06}) and 050724 (light grey circles;
\citealt{gbp+06}). The data for GRBs\,051221A and 050724 have been
scaled for clarity. Grey lines trace the afterglow evolution with a
break for GRB\,051221A at $\approx 5$ days and no break for
GRB\,050724A to $\approx 22$ days.
\label{fig:xraylc}} 
\end{figure}

The temporal behavior of the X-ray afterglow flux is characterized by
a steady power-law decline until $\delta t \approx 2$ days when there is a
significant steepening in the light curve (Figure~\ref{fig:xraylc}). A
single power law model with a decline rate determined by the X-ray
data at early times ($t \lesssim 2$ days) provides a poor fit to the
late-time data (dotted line in Figure~\ref{fig:xraylc}); in particular, it
overestimates the {\it Chandra} detection and upper limit. To
quantitatively assess the shape of the X-ray light curve, we therefore
invoke a broken power law model, given by

\begin{equation}
F_X=F_{X,0}\left[\left(\frac{t}{t_j}\right)^{\alpha_{X,1}s}+\left(\frac{t}{t_j}\right)^{\alpha_{X,2}s}\right]^{1/s},
\label{eqn:bpl}
\end{equation}

\noindent where $F_{X,0}=2^{1/s}F_X(t=t_j)$, $\alpha_{X,1}$ and
$\alpha_{X,2}$ are the power law indices pre- and post-break,
respectively, $t_j$ is the break time in seconds, and $s$ is a
dimensionless smoothness parameter that characterizes the sharpness of
the break.  We perform a three-parameter $\chi^2$-grid search over
$F_{X,0}$, $\alpha_{X,1}$ and $t_j$. If we use a relatively sharp
break (e.g. $s \approx -10$), the {\it Chandra} $3\sigma$ upper limit
constrains $\alpha_{X,2}\lesssim -1.7$.  If we allow for a smoother
break (e.g. $s \approx -1$), $\alpha_{X,1}$ remains unchanged but the
break occurs at later times ($t_j \approx 4$ days) and $\alpha_{X,2}$
is required to have a steeper value of $\lesssim -2.2$ to accommodate
the \chandra\ upper limit. This scenario generally provides a poorer
fit to the last \chandra\ and \swift/XRT points. We therefore adopt
the sharp-break scenario. Fixing $s=-10$ and $\alpha_{X,2}=-2.1$, we
find a best-fit broken power law model characterized by
$F_X(t_j)=(1.36 \pm 0.45) \times 10^{-13}$ erg cm$^{-2}$ s$^{-1}$,
$\alpha_{X,1}=-0.78 \pm 0.05$, and $t_j=2.0 \pm 0.5$ days ($1\sigma$,
$\chi^2_\nu=1.1$ with $15$ d.o.f.; Figure~\ref{fig:xraylc}). This
best-fit model is shown in Figure~\ref{fig:xraylc}. The best fit
parameters are independent of our choice of $\alpha_{X,2}$ between
$-1.7$ and $-3$. We also note the presence of a slight flux
enhancement relative to the power law decay at $\delta t \approx 3
\times 10^{4}$ s ($0.35$ days). If we remove these points
from our fits, the resulting best-fit parameters remain unaffected.

The required change in the temporal index is $\Delta\alpha_{12}
\gtrsim 0.9$. There are several possibilities that can explain breaks
in GRB afterglow light curves. The first scenario is the transition of
the cooling frequency across the band, but this only predicts
$\Delta\alpha=0.25$ \citep{spn98}. An alternative possibility is the
cessation of energy injection, either from refreshed shocks or a
long-lasting central engine (e.g., \citealt{rm98,sm00,zm02}), which
has been used to explain the termination of a shallow decay or plateau
phase in the X-ray and optical light curves of several long
GRBs. However, these cases all exhibit earlier temporal breaks at
$\sim 10^3-10^4$ sec with $\Delta\alpha_{12} \sim 0.7$ ($\alpha_{X,1}
\approx -0.5$, $\alpha_{X,2} \approx -1.2$;
\citealt{nkg+06,zfd+06,lzz07}). Attributing the break in \grb\ to the
cessation of central engine activity would require sustained energy
injection from the start of XRT monitoring to the break time, $\sim
100$ seconds to $2$ days, whereas the timescales of energy injection
for long GRBs are $\lesssim$ few hours
\citep{nkg+06,zfd+06,lzz07,rlb+09}. Single episodes of energy
injection have also been observed in two short GRBs: 051221A and
050724A \citep{bpc+05,sbk+06,bgc+06,gbp+06}). The light curve of
GRB\,051221A, which exhibits a power law decay with index
$\alpha_{X,1}=-1.1$, a plateau, and a return to the same power law
($\Delta\alpha_{12}=0$), is interpreted as a single period of energy
injection \citep{sbk+06,bgc+06}. A super-imposed flare on the light
curve of GRB\,050724A with a single underlying decay index of
$\alpha_{X,1}=-0.98$ is also possibly related to late-time
reactivation of the central engine (\citealt{bpc+05,gbp+06};
Figure~\ref{fig:xraylc}). Neither of these light curves resemble the
behavior of \grb, where the change in slope is substantially greater.

Yet another possibility to explain the break is a sharp change in the
external density. However, models for density jumps in a uniform
medium \citep{ng07} predict that the density would need to decrease by
greater than a factor of $\sim 10^3$ to account for the observed
$\Delta\alpha_{12}>0.9$ steepening. More realistic density contrasts
of $\sim 10$ predict $\Delta\alpha_{\rm max} \approx 0.4$ in optical
and X-ray afterglow light curves \citep{ng07}.

Finally, we consider that the observed steepening is a jet break, when
the edge of a relativistically-beamed outflow becomes visible to the
observer and the jet spreads laterally \citep{sph99,rho99}. This model
is often adopted to explain $\Delta\alpha_{12} \sim 1$ in the light
curves of long GRBs (e.g., \citealt{fks+01,bfk03,rlb+09}) and has been
observed in one other short burst, GRB\,051221A ($\Delta\alpha_{12} \sim
0.9$, Figure~\ref{fig:xraylc}; \citealt{sbk+06,bgc+06}). Given the
similarity in $\Delta\alpha_{12}$ and the timescales of jet breaks in
both short and long GRBs, we conclude that the observed steepening in
the light curve of \grb\ is best explained by a jet break at $t_j=2.0
\pm 0.5$ days.

\subsection{Afterglow Properties}

We utilize our radio, optical and X-ray observations to constrain the
explosion properties and circumburst environment of \grb. In
particular, we adopt the standard synchrotron model for GRB afterglows
\citep{spn98,gs02} which provides a mapping from observable properties
to the isotropic-equivalent kinetic energy ($E_{\rm K,iso}$), circumburst density ($n_0$),
and the fractions of post-shock energy in radiating electrons ($\epsilon_e$) and
magnetic fields ($\epsilon_B$). We use data at the time of the radio
and first optical observations ($\delta t = 17.7$ hours), as well as
the decay indices from the full X-ray light curve.

First, we constrain the electron power-law index $p$, using a
combination of temporal and spectral information. From the X-ray light
curve, we measure $\alpha_{X,2} \lesssim -1.7$
(Section~\ref{sec:jb}). For $p=-\alpha_{X,2}$, appropriate for a
spreading jet \citep{sph99}, we can then constrain $p \gtrsim 1.7$. To
further constrain $p$ and investigate the location of the cooling
frequency, $\nu_c$, we compare the values $\alpha_{X,1}=-0.78 \pm
0.05$ and $\beta_{X}=-1.04 \pm 0.16$ ($\beta_X=1-\Gamma$; $1\sigma$) to the closure relations for a relativistic
blastwave in a constant density ISM-like medium for $p>2$, a typical
environment expected for a short GRB from a non-massive star
progenitor \citep{sph99,gs02}. If $\nu_c > \nu_X$ then the
independently-derived values for $p$ from the temporal and spectral
indices are inconsistent: $p=2.0 \pm 0.07$ from $\alpha_{X,1}$, and
$p=3.1 \pm 0.32$ from $\beta_X$ (errors are $1\sigma$).

However, if $\nu_c < \nu_X$ we obtain $p=1.7 \pm 0.07$ from
$\alpha_{X,1}$, \citep{gs02} which is consistent with the $p$ value
inferred from $\alpha_{X,2}$, but yields a divergent total integrated
energy in electrons unless a break at high energies in the
distribution is invoked. Although a flat electron distribution ($p<2$)
is possible and not uncommon (e.g. \citealt{dc01,pk01,rlb+09}), the
standard relations for $1<p<2$ yield $p=0.84 \pm 0.25$ from
$\alpha_{X,1}$. This solution is not self-consistent, and would also
require an unusually sharp break of $\Delta p \gtrsim 1.2$ in the
electron distribution. Continuing with the assumptions that $\nu_c <
\nu_X$ and $p>2$, we obtain $p=2.1 \pm 0.32$ from $\beta_X$, which is
marginally consistent with the value inferred from the temporal index. Put
another way, $\alpha-3\beta/2 = 0.77 \pm 0.30$, which satisfies the
closure relation for $\nu_c<\nu_X$ \citep{spn98}. We therefore
conclude that $\nu_c<\nu_X$. We note that the spectral index is
generally more reliable in the determination of $p$ because it remains
constant over time and is not subject to alternative processes such as
energy injection or flaring. In this case, the same $\beta_X$ was also
independently determined from both the XMM and XRT data sets
(Table~\ref{tab:xrayspec}). Therefore for the rest of our
calculations, we take a reasonable value of $p=2.1$ as determined from
$\beta_X$.

We next determine a set of constraints on $n_0$ and $E_{\rm K,iso}$
based on the X-ray flux density, radio limit, and the condition that
$\nu_c<\nu_X$. First, we use the X-ray afterglow emission as a proxy
for $E_{\rm K,iso}$ assuming the X-ray emission is from the forward
shock. For $\nu_c<\nu_X$ at the time of our broad-band observations
($\delta t=17.7$ hours), we use $F_X=0.032\,\mu$Jy at $\nu_{X}=2.4
\times 10^{17}$ Hz ($1$ keV), and $p=2.1$ to obtain \citep{gs02}

\begin{equation}
E_{\rm K,iso} \approx 2.2 \times 10^{50} (1+z)^{-1} \epsilon_e^{-1.07} \epsilon_B^{-0.024} d_{\rm{L,28}}^{1.95}\,\,\,{\rm erg}
\label{eqn:E_K}
\end{equation}

\noindent where $d_{\rm L,28}$ is the luminosity distance in units of
$10^{28}$ cm. Next, we use $E_{\rm K,iso}$ to constrain $n_0$. Using our 3$\sigma$
EVLA limit of $F_{\nu,{\rm rad}} \lesssim 39$ $\mu$Jy, we can determine an upper limit on
$n_0$ under the reasonable assumption that our observed radio band
($\nu=5.8$ GHz) is above the self-absorption frequency
($\nu_{a}<\nu_{\rm rad} <\nu_m$; $F_{\nu, {\rm rad}} \propto n_{0}^{1/2}$) at the time of
observations. For this scenario \citep{gs02},

\begin{equation}
 n_0 \lesssim 1.7 \times 10^{-3} E_{\rm K,iso,52}^{-5/3} (1+z)^{-5/3} d_{L,28}^4 \epsilon_e^{4/3} \epsilon_B^{-2/3}\,\,\,{\rm cm^{-3}},
\label{eqn:nup}
\end{equation}

\noindent where $E_{\rm K,iso,52}$ is in units of $10^{52}$ erg and
$n_0$ is in cm$^{-3}$. As noted in Section~\ref{sec:evla}, if the
marginal radio detection is indeed real, then this upper bound can be
replaced with an equality. Finally, we can place a lower limit on the
density using the condition that $\nu_c<\nu_X$ (i.e., $\nu_c \lesssim
2.4 \times 10^{16}$ Hz; $0.1$ keV)

\begin{equation}
n_0 \gtrsim 4.5 \times 10^{-4} (1+z)^{-1/2} \epsilon_B^{-3/2} E_{\rm K,iso,52}^{-1/2}\,\,\,{\rm cm^{-3}}.
\label{eqn:nlow}
\end{equation}

\begin{figure}
\centering \plottwo{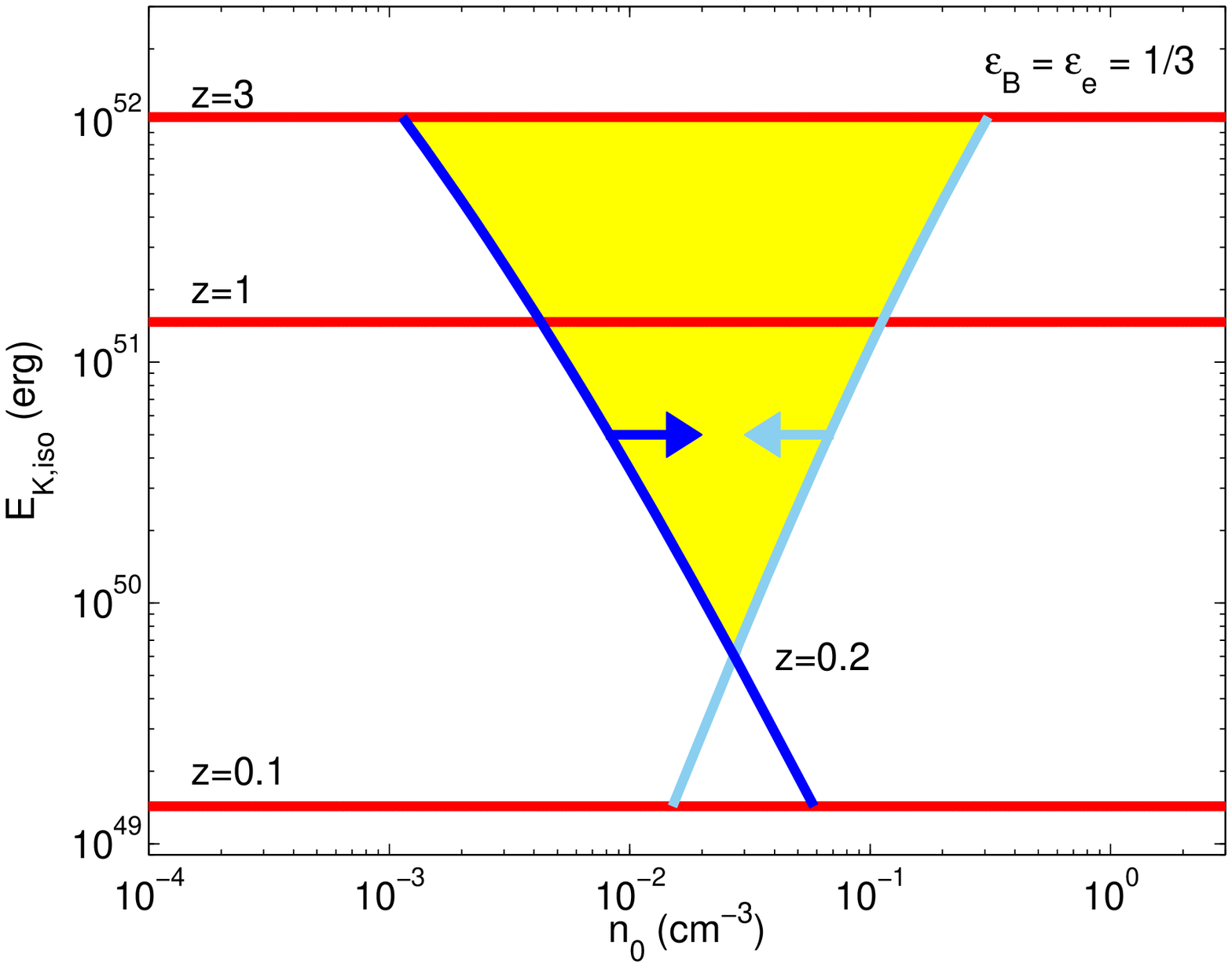}{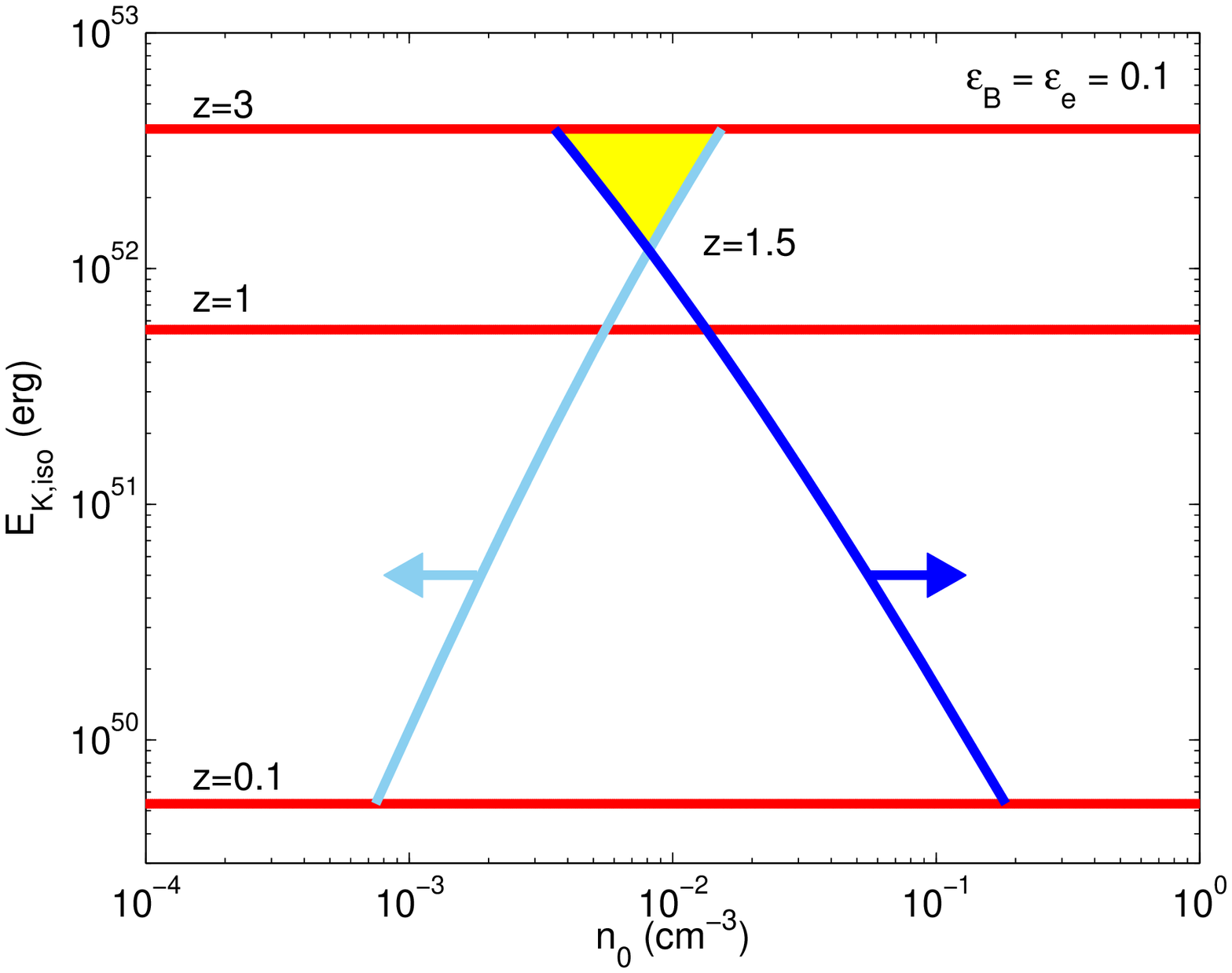}
\caption{Isotropic-equivalent kinetic energy and circumburst density
parameter space for \grb\ assuming $\epsilon_e=\epsilon_B = 1/3$
(left) and $\epsilon_e=\epsilon_B = 0.1$ (right). The lower limit on
the density (dark blue) is set by the condition that $\nu_c<\nu_X$
(Equation~\ref{eqn:nlow}) while the upper limit (light blue) is set by
radio observations (Equation~\ref{eqn:nup}). Also plotted are the
values for $E_{\rm K,iso}$ at $z=0.1,1$ and $3$ (red). The allowable
parameter space set by these constraints is filled in yellow.
\label{fig:EKn}} 
\end{figure}

\noindent It is clear that $E_{\rm K,iso}$ and $n_0$ depend
sensitively on our choice of $z$, $\epsilon_e$ and $\epsilon_B$. The
fractions $\epsilon_e,\epsilon_B$ are not expected to exceed $\sim
1/3$. We therefore calculate $E_{\rm K,iso}$ for two cases: I:
$\epsilon_e=\epsilon_B = 1/3$, and II: more typical values of
$\epsilon_e=\epsilon_B=0.1$. We then calculate the range of allowed
$n_0$ set by Equations~\ref{eqn:nup} and \ref{eqn:nlow}, which becomes
wider as the redshift increases. For Case~I, this requires that $z
\gtrsim 0.2$, below which the constraints on $n_0$ conflict
(Figure~\ref{fig:EKn}). At the median observed redshift of the short
GRB population, $z \sim 0.5$, we obtain $E_{\rm K,iso} \approx 3.7
\times 10^{50}$ erg and $n_0=0.01-0.06$ cm$^{-3}$. For Case~II, the
constraints on $n_0$ require a higher redshift of $z \gtrsim 1.5$
(Figure~\ref{fig:EKn}). For a fiducial redshift of $z=1.5$, we obtain
$E_{\rm K,iso} \approx 1.2 \times 10^{52}$ erg and $n_0=0.008$
cm$^{-3}$. The parameters for the two cases are listed in
Table~\ref{tab:param}. Although we cannot distinguish between these
two scenarios, both cases require low circumburst densities of $n \sim
0.01$ cm$^{-3}$.

\subsection{Jet Opening Angle}
\label{sec:angle}

In the context of a jet break, we use the time of the break from the
X-ray light curve ($2.0 \pm 0.5$ days) and the circumburst density and
energy estimates from the previous section to constrain $\theta_j$.
The time of the break is a direct reflection of the jet opening angle
\citep{sph99,fks+01},

\begin{equation}
\theta_j=0.1 t_{j,{\rm d}}^{3/8}(1+z)^{-3/8}E_{\rm K,iso,52}^{-1/8}n_{0}^{1/8}
\label{eqn:jb}
\end{equation}

\noindent where $t_{j,{\rm d}}$ is expressed in days. For our
fiducial Case~I ($z=0.5, \epsilon_e=\epsilon_B=1/3$), $E_{\rm
K,iso} \approx 3.7 \times 10^{50}$ erg and $n \approx 0.01-0.06$ cm$^{-3}$
give $\theta_j=7-8^{\circ}$. This leads to a beaming correction on the
energy of $f_b \equiv [1-{\rm cos}(\theta_j)]=0.007-0.01$, and
therefore a true kinetic energy $E_K=f_b E_{\rm K,iso} \approx (3-4)
\times 10^{48}$ erg (Table~\ref{tab:param}). To estimate the
beaming-corrected $\gamma$-ray energy, we infer $E_{\gamma,{\rm iso}}$
from the \swift/BAT fluence and apply a bolometric correction factor
of 5 to roughly convert to a representative observed $\gamma$-ray
energy range of $\sim 10-1000$ keV.  This factor is derived from short
GRBs observed by satellites with wider energy coverage
\citep{ber10,mzb+12}. We obtain $E_{\gamma, {\rm iso}}=2.1 \times
10^{50}$ erg and therefore a true $\gamma$-ray energy of $E_{\gamma}
\approx 2 \times 10^{48}$ erg.

For Case~II ($z=1.5, \epsilon_e=\epsilon_B=0.1$), where $n_0 \approx
0.008$ cm$^{-3}$ and $E_{\rm K,iso} \approx 1.2 \times 10^{52}$ erg,
we obtain a smaller opening angle of $\theta_j \approx 3^{\circ}$. This leads to $f_b
\approx 1.4 \times 10^{-3}$ and hence, $E_{\gamma} \approx 3 \times 10^{48}$ erg and $E_{K}
\approx 2 \times 10^{49}$ erg.

In both cases, the true $\gamma$-ray energy is few $\times 10^{48}$ erg
while the kinetic energy is an order of magnitude higher at $z=1.5$
than at $z=0.5$. This results in a {\it lower} $\gamma$-ray conversion
efficiency ($\eta_\gamma \equiv E_{\gamma}/E_{\rm tot}$) for Case~II
of $0.15$ compared to $0.3-0.4$ for Case~I (Table~\ref{tab:param}). The
total energy even for Case~II is $\sim 10-100$ times lower that
for long GRBs.

\subsection{Extinction}

We investigate the presence of extinction by comparing the unabsorbed
X-ray flux and the optical non-detection at $\delta t=17.7$
hour. Since we do not know the exact location of the cooling
frequency, we assume a maximum value $\nu_{c,{\rm max}}$ of $2.4
\times 10^{17}$ Hz ($1$ keV) and extrapolate the X-ray flux to the
optical band using the shallowest possible slope of
$\beta=-(p-1)/2=-0.55$ to obtain the lowest bound on the expected
optical afterglow flux in the absence of extinction; any other
assumption for the location of $\nu_c<\nu_X$ would result in a higher
predicted optical flux density. For $p=2.1$ we estimate $F_{\nu,{\rm
opt}}\approx 1.1$ $\mu$Jy ($i=23.8$ mag). Given that our observed
$3\sigma$ upper limit is $i \gtrsim 24.4$ mag, this implies a lower
limit on the optical extinction in excess of the Galactic value of
$A_i \gtrsim 0.6$ mag\footnotemark\footnotetext{We note that for $p
\lesssim 1.9$, no host galaxy extinction is required}. In the rest
frame of the burst for a Milky Way extinction curve, this translates
to $A_V^{\rm host} \gtrsim 0.6$ mag for $z=0.5$ and $A_V^{\rm host}
\gtrsim 0.2$ at $z=1.5$ \citep{ccm89}. Using Galactic relations
between $N_H$ and $A_V$, $N_{\rm H,int}/A_V \approx (1.7-2.2) \times
10^{21}$ \citep{ps95,wat11}, we infer lower limits of $N_{\rm H,int}
\gtrsim 10^{21}$ cm$^{-2}$ at $z=0.5$ and $N_{\rm H,int} \gtrsim 4.4
\times 10^{20}$ cm$^{-2}$ at $z=1.5$, consistent with our value of
$7.5 \times 10^{21}$ cm$^{-2}$ ($z=0$) derived from the X-ray spectrum
(Table~\ref{tab:xrayspec}). However, an appreciable extinction is
unexpected given the burst's location on the outskirts of its
potential host galaxy. We note that the burst is located at Galactic
coordinates ($l,b$)$=(359.3^{\circ},-19.4^{\circ})$ which is toward
the Galactic Bulge on a steep gradient in the dust map \citep{sfd98}
and thus may be subject to substantial ($\sim 30\%$) uncertainties in
the Galactic extinction\footnotemark\footnotetext{Using a
high-resolution ($\theta_{\rm FWHM}=15''$) WISE $12\,\mu$m map, we do
not see strong evidence for any thin dust filaments at the location of
the burst which would result in $>30\%$ uncertainties in the Galactic
$A_V$ (D. Finkbeiner, private comm.)}. Taking this uncertainty into
account reduces the required $A_V^{\rm host}$ to $\gtrsim 0.2-0.3$ mag
depending on the redshift of the burst. \\

\section{Discussion}

\subsection{Environment}

From our broad-band observations, we constrain the circumburst density
of \grb\ to $n_0 \sim 0.01$ cm$^{-3}$ which is consistent with
the low values inferred for a few previous short GRBs
\citep{sbk+06,pan06,sdp+07,pmg+09,ber10,fbc+11}. The inferred density
fits well with the framework of NS-NS/NS-BH binary progenitor systems,
which may be subject to substantial kicks from their host galaxies and
are predicted to typically occur at densities of $\sim\!10^{-6}-1$
cm$^{-3}$ \citep{pb02,bpb+06}.

\grb\ has an offset of $\approx\!0.80''$ from its most probable host
galaxy (G3; Figure~\ref{fig:gmos_psf_panel}). For redshifts between
$z=0.5-1.5$, this translates to a projected physical offset of
$\approx 5-7$ kpc, which is comparable to the median value of $\sim 5$
kpc for well-localized short GRBs with host associations
\citep{fbf10,cld+11}. Although G3 has the lowest probability of chance
coincidence by an order of magnitude (Figure~\ref{fig:111020a_cc}), we
cannot rule out the possibility that G3 is a faint star. The next most
probable galaxies, G1 and G2, are situated $2.8''$ ($17-24$ kpc) and
$6.5''$ ($40-56$ kpc), respectively, from \grb\
(Figure~\ref{fig:gmos_psf_panel}). If the burst originated from one of
these galaxies, this would put \grb\ at the high end of the observed
offset distribution, similar to the growing sub-class of apparently
``hostless'' short GRBs which likely occur $\gtrsim 30$ kpc from their
host galaxies \citep{ber10}. All of these inferred offsets are
consistent with predicted offset distributions of NS-NS/NS-BH binaries
originating in Milky Way-type galaxies
\citep{bsp99,fwh99,bpb+06,sdc+10}.

Most short GRB host galaxies with confirmed spectroscopic redshifts
have measured luminosities of $L_{B} \approx 0.1-1 L_*$
\citep{bfp+07}. The apparent magnitude of G3 is $i \approx 24.3$,
which corresponds to $L_B \approx 0.1-1 L_*$ over $z \approx 0.5-2.3$
when compared to the luminosity function of galaxies at corresponding
redshifts in the DEEP2 and LBG surveys \citep{wfk+06,rs09}. This is
consistent with the redshift range inferred from the afterglow.

We next investigate the nature of the dust and gas in the environment
of \grb\ through an analysis of $A_V^{\rm host}$ and $N_{\rm
H,int}$. We have shown that the burst requires dust extinction of
$A_V^{\rm host} \gtrsim 0.2-0.6$ mag, depending on the redshift of the
burst and the uncertainty in Galactic extinction. We have also
measured a neutral Hydrogen column density intrinsic to the burst
environment of $N_{\rm H,int}=(7.5 \pm 2.0) \times 10^{21}$ cm$^{-2}$
at $z=0$, which becomes higher for any other choice of $z$. High
values of both dust extinction and X-ray absorption have been linked
to ``dark'' GRBs (e.g. \citealt{pcb+09,csm+11}) which have optically
sub-luminous afterglows compared to their X-ray or NIR counterparts
and can quantitatively be classified by $|\beta_{OX}| \lesssim
|\beta_X|-0.5$ (\citealt{vkg+09}; see also \citealt{jhf+04}). With
$|\beta_X|=1.0$ and $|\beta_{OX}| \lesssim 0.46$, \grb\ is consistent
with this definition of dark GRBs. While optical extinction intrinsic
to long GRB environments is not uncommon and commensurate with their
origin in dusty, star-forming regions, evidence for substantial
extinction has been reported for only one other short burst,
GRB\,070724A, which required $A_V^{\rm host} \gtrsim 2$ mag to explain
the suppression of optical emission relative to the NIR
\citep{bcf+09,ktr+10}. The location of GRB\,070724A on the outskirts
of its host galaxy, $\sim 5$ kpc from the center, suggested either an
origin in a star-forming region or a progenitor system which produced
the dust itself \citep{bcf+09}. The potentially appreciable extinction
and the location with respect to its putative host suggests that the
same conclusions may be drawn for \grb.

\begin{figure}
\centering
\includegraphics[angle=0,width=6.4in]{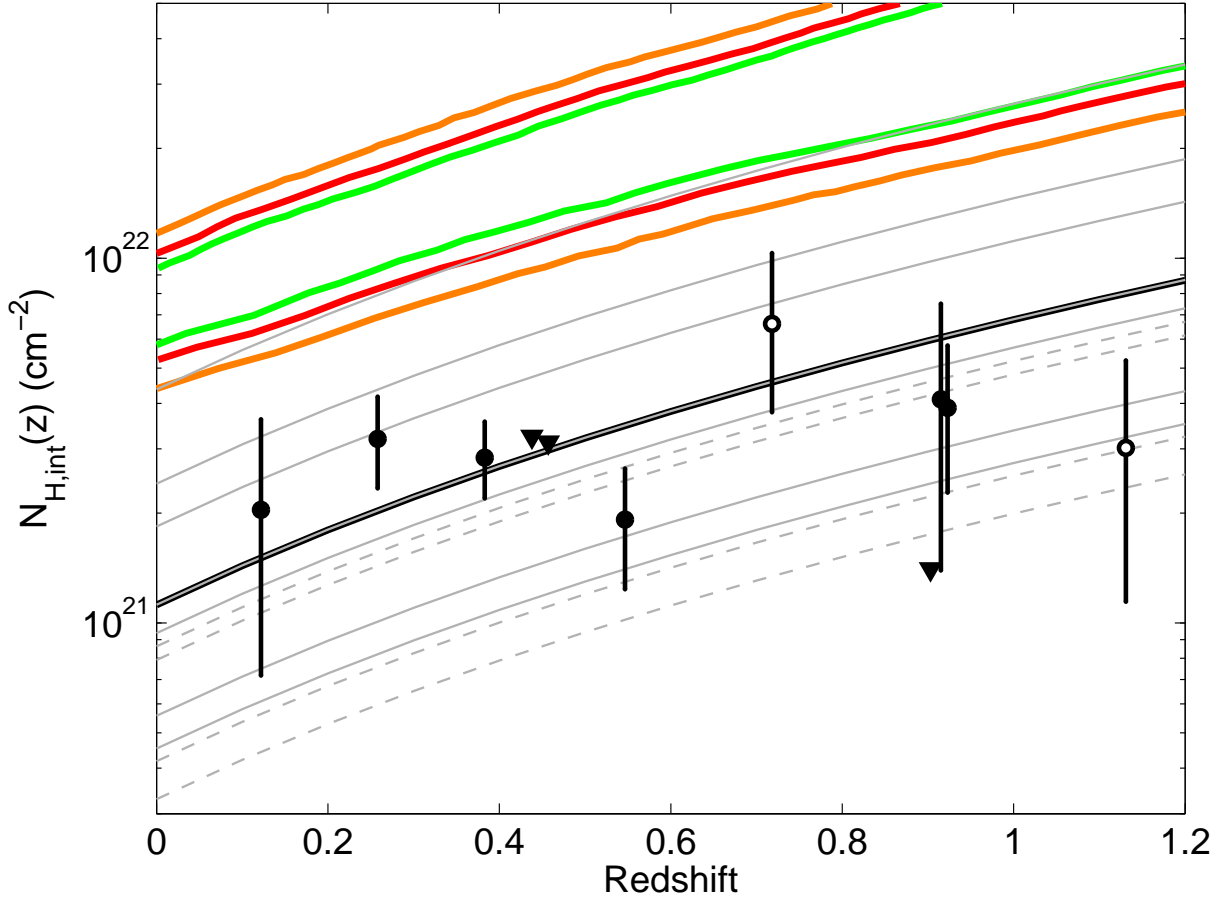}
\caption{Excess neutral hydrogen column density, $N_{\rm H,int}$, versus redshift for
\grb\ ($1$, $2$ and $3\sigma$ intervals denoted by green, red and
orange lines) along with $6$ \swift\ short GRBs with measured
redshifts and optical afterglows (black filled circles) and $2$
(GRBs\,060801 and 101219A) with only X-ray afterglows (open
circles). Also plotted are $11$ short GRBs without secure redshifts
(grey lines), $4$ of which have only upper limits on $N_{\rm H,int}$
(grey dashed). For GRBs without redshifts, the $N_{\rm H,int}$ value
at $z=0$ is scaled by $(1+z)^{2.6}$ \citep{gw01}. Errors and upper
limits are at the $90\%$ confidence level. The weighted mean for all
short GRBs (black line) over the redshift interval $z=0-1.2$ is also
shown. \grb\ has the highest $N_{\rm H,int}$ of a short GRB to date
and is well above the mean for short GRBs.
\label{fig:NHz}} 
\end{figure}

On the other hand, the relation between $N_{\rm H,int}$ and the
darkness of a burst is less clear. A recent study of long dark GRBs
shows them to have higher intrinsic column densities than non-dark
GRBs, which suggests that the darkness of a burst is largely due to
absorption by circumburst material \citep{csm+11}. To investigate this
relationship for \grb, we extract spectra and best-fitting $N_{\rm
H,int}$ for all short GRBs with XRT-detected afterglows in the same
manner as \grb\ (see Section~\ref{sec:specfit}), over time ranges with
no evidence for spectral evolution. There are $22$ short bursts with
sufficient X-ray counts to perform spectral analysis, $11$ of which
have known redshifts (Table~\ref{tab:NH}). We find a short GRB
weighted average of $N_{\rm {H,int}}$($z=0$)$=(1.1 \pm 0.14) \times
10^{21}$ cm$^{-2}$ ($90\%$ c.l.; Figure~\ref{fig:NHz}). In comparison,
\grb\ has a high value of $N_{\rm H,int}=(7.5 \pm 2.0) \times 10^{21}$
cm$^{-2}$ at $z=0$ (Figure~\ref{fig:NHz}). Taken at face value, it is
surprising to find a large $N_{\rm H,int}$ for a substantial offset, and may
suggest that the burst occurred in a relatively metal-rich
environment.

\subsection{Beaming, Energetics, and Rates}
\label{sec:ber}

\begin{figure}
\centering
\includegraphics[angle=0,width=6.4in]{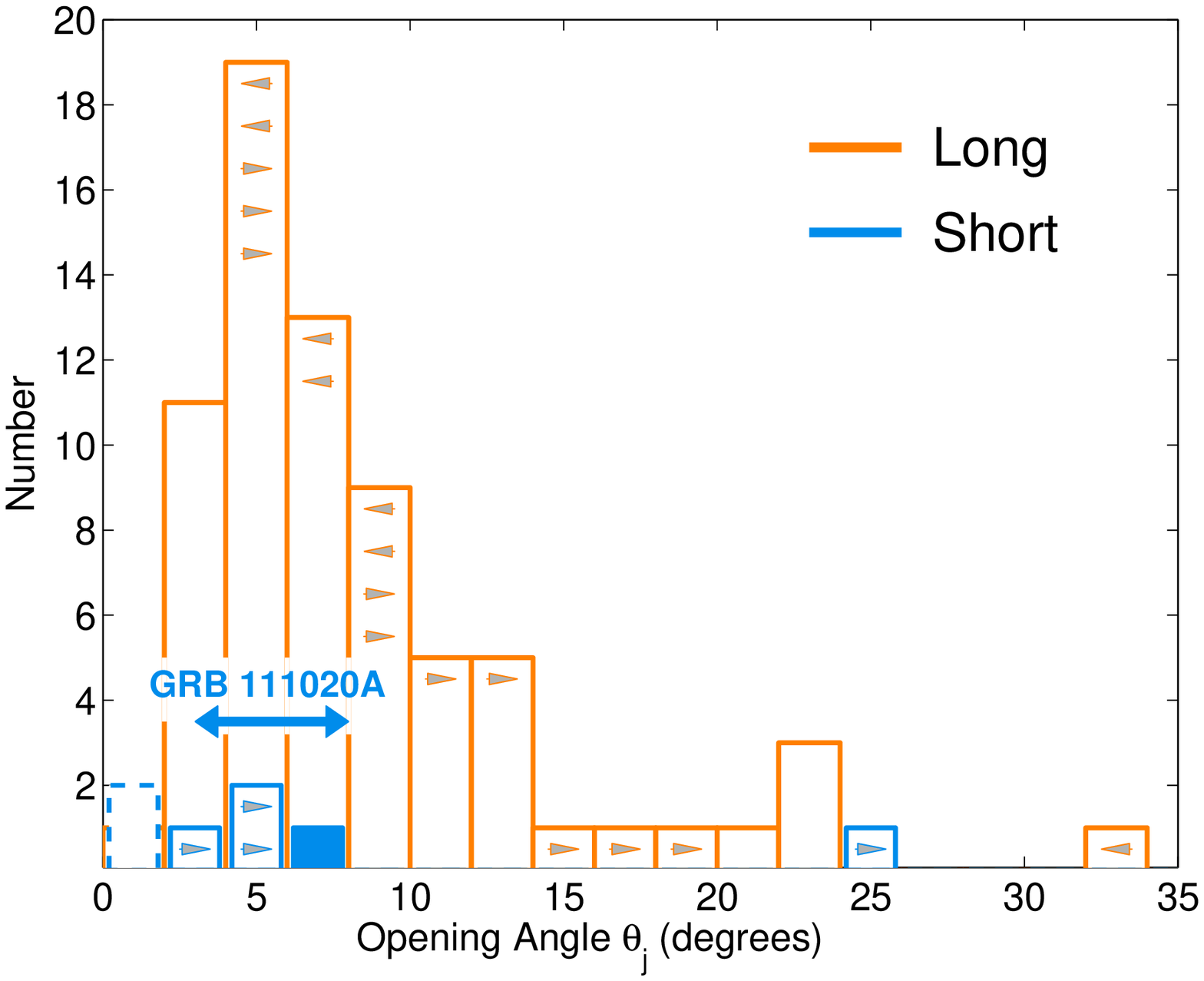}
\caption{Distribution of opening angles for long (orange) and short
(blue) GRBs. Arrows represent upper and lower limits. The long GRB
population includes pre-\swift\ \citep{fks+01,bkf03,bfk03,ggl04,fb05},
\swift\ \citep{rlb+09,fkg+11}, and {\it Fermi}
\citep{cfh+10,gpb+11,cfh+11} bursts. The opening angle for \grb\
ranges from $\sim 3-8^{\circ}$ (depending on the redshift), while
GRB\,051221A has $\theta_j \approx 7^{\circ}$
\citep{sbk+06,bgc+06}. Tentative jet breaks (blue dashed) for
GRBs\,061201 \citep{sdp+07} and 090510 \citep{dsk+10,nkk+12} are at
$\sim 1^{\circ}$. Short GRB lower limits are from the non-detection of
jet breaks in \swift/XRT data (this work, revised from
\citealt{chp+12}) and {\it Chandra} data for GRB\,050724A
\citep{bpc+05,gbp+06}.
\label{fig:angle}} 
\end{figure}

We uncover a break in the X-ray light curve of \grb\ at $\approx\!2$
days, which we interpret as a jet break
(Section~\ref{sec:jb}). Depending on our values for $z$, $\epsilon_e$
and $\epsilon_B$, we infer an opening angle of $\approx
3-8^{\circ}$. This is reminiscent of the first jet break discovery in
GRB\,051221A, with $\theta_j \approx 7^{\circ}$ \citep{sbk+06,bgc+06},
and suggests that at least a fraction of these events are highly
collimated. In addition, temporal breaks at $t_j \lesssim$ few hours
in GRBs\,061201 \citep{sdp+07} and 090510 \citep{dsk+10,nkk+12}, if
interpreted as jet breaks, lead to $\theta_j \approx 1^{\circ}$
(Figure~\ref{fig:angle}). However, these two cases resemble early
breaks in long GRBs that are generally attributed to the cessation of
energy injection, and not collimation.

Although the remaining short GRB afterglow data is sparse, the lack of
observed jet breaks in their X-ray light curves can be used to place
lower limits on the opening angles. Indeed, \chandra\ observations of
GRB\,050724A out to $22$ days indicated $\theta_j \gtrsim 25^{\circ}$,
consistent with a spherical explosion \citep{gbp+06}. A recent study
by \citet{chp+12} analyzed the sample of short GRB \swift/XRT light
curves up to August 2011 with monitoring $\gtrsim 1$ day which
included $6$ additional events, and inferred $\theta_j \gtrsim
6-16^{\circ}$, assuming $n_0=1$ cm$^{-3}$ for all bursts. We revise
this analysis for 3 of the events with robust X-ray light curves
(GRBs\,070714B, 070724A, 071227; data analysis prescriptions from
\citealt{mzb+12}) employing a more representative $n_0 \approx
10^{-2}$ cm$^{-3}$ (e.g. \citealt{sbk+06} and this work).  We derive
$E_{\gamma,{\rm iso}}$ from the reported fluences, applying a
bolometric correction when necessary to represent an energy range of
$\sim 10-1000$ keV, and infer more realistic lower limits of $\gtrsim
2-6^{\circ}$ (Figure~\ref{fig:angle}). These limits are indeed lower
than the detected values for GRBs\,051221A and 111020A, and therefore
do not add strong constraints on the distribution. We caution that the
sample presented here represents only the $\sim 30\%$ of the \swift\
short GRB population that have bright X-ray afterglows and relatively
slow flux decline rates; the remaining fraction do not have detectable
X-ray afterglows or fade too quickly so constraints cannot be placed
on their collimation.

There are now two short GRBs with opening angle measurements, two with
measurements based on more tentative early breaks, and an additional
four with lower limits (Figure~\ref{fig:angle}). These early
constraints create a distribution that may mimic the distribution for
long GRBs, which ranges from $\sim\!2-20^\circ$ with a median of
$7^{\circ}$ (Figure~\ref{fig:angle};
\citealt{fks+01,bkf03,bfk03,ggl04,fb05,rlb+09,cfh+10,fkg+11,gpb+11,cfh+11}). More
events are needed to assess the real differences between the
distributions. However, simulations of post-merger black hole
accretion predict jets with $\theta_j \sim 5-20^{\circ}$
\citep{ajm+05,ros05,rgb+11} to several tens of degrees
\citep{rj99a,rgb+11} depending on the mechanism of energy extraction
and Lorentz factor, so there are expectations on theoretical grounds
that the short GRB distribution is wider.

The first major ramification of collimation is the correction to the
total energy release: the true energy is lower than the
isotropic-equivalent value by the beaming factor, $f_b$.  For \grb\
with an opening angle of $\approx 3-8^{\circ}$, this correction factor is
substantial, $0.001-0.01$. Depending on the redshift, the
beaming-corrected energy of \grb\ is $E_{\gamma} \approx (2-3)
\times\!10^{48}$ erg (Table~\ref{tab:param}) which is an order of
magnitude lower than for GRB\,051221A with $E_{\gamma}\!\approx\!(1-2)
\times 10^{49}$ erg \citep{sbk+06,bgc+06} and GRB\,050724A with
$E_{\gamma} \approx (0.4-4) \times 10^{50}$ erg \citep{gbp+06}. The three
remaining events with opening angle lower limits, GRBs\,070714B,
070724A and 071227, have ranges of $E_{\gamma} \approx
10^{48}-10^{51}$ erg, where the upper bound is set by the
isotropic-equivalent $\gamma$-ray energy in the $\approx 10-1000$ keV
band. The small population of short GRBs with measured $E_{\gamma}$
therefore has a median value of $E_{\gamma}\!\sim\!10^{49}$ erg,
which is an order of magnitude below \swift\ long GRBs
\citep{kb08,rlb+09} and $2$ orders of magnitude below the pre-\swift\
population \citep{fks+01,bfk03}. Again, this sample is incomplete
because we can only measure $E_{\gamma}$ for bursts with
well-constrained opening angles.

In a similar vein, we compare the beaming-corrected kinetic energy and
total energy ($E_K$, $E_{\rm tot}$) of \grb\ to the values for other
short bursts. Because $E_{\rm K,iso}$ is more sensitive to our choices
for $z$, $\epsilon_e$ and $\epsilon_B$, we infer different values for
Case I and II. For Case I, we infer $E_K \approx (3-4) \times 10^{48}$
erg, $E_{\rm tot}=E_{\gamma}+E_K \approx (5-6) \times 10^{48}$ erg,
and $\eta_\gamma \approx 0.3-0.4$. For Case II, we calculate $E_K
\approx 2 \times 10^{49}$ erg, $E_{\rm tot} \approx 2 \times 10^{49}$
erg and $\eta_\gamma \approx 0.15$. GRB\,051221A had $E_K \approx 8 \times
10^{48}$ erg and a total energy release of $\approx 2.5 \times 10^{49}$ erg
\citep{sbk+06,bgc+06} while GRB\,050724 had a total energy of
$10^{50}-10^{51}$ erg. With $E_{\rm tot} \approx (0.5-2) \times
10^{49}$ erg, \grb\ may be on the low end of the total energy
distribution, but more events with beaming-corrected
energies are needed to better characterize the distribution for short
GRBs.

The true total energy release of short GRBs has strong implications on
the energy extraction mechanism. Two primary mechanisms, the thermal
energy release from $\nu\bar{\nu}$ annihilation in a baryonic outflow
\citep{jar93,mhi+93} and magnetohydrodynamic (MHD) processes in the
black hole's accretion remnant (e.g. \citealt{bz77,rrd03}), give
different estimates for the expected energy release. Predictions for
$\nu\bar{\nu}$ annihilation are largely dependent on the mass of the
disc and efficiency to produce pairs. Simulations of an outflow due to
$\nu\bar{\nu}$ annihilation suggest beaming-corrected total energy
releases could reach $10^{48}-10^{49}$ erg
\citep{rj99a,rj99b,pwf99,ros05,baj+07,lr07}. Higher energy releases
can be obtained from MHD processes, which can produce luminosities of
$\gtrsim 10^{52}$ erg s$^{-1}$ ($\gtrsim 10^{50}$ erg s$^{-1}$ when
corrected for beaming; \citealt{pwf99,rrd03,lr07}) depending on the
nature of the magnetic field amplification. While the true energy
releases of GRBs\,051221A and 050724A suggest that MHD processes may
be powering these events \citep{bpc+05,gbp+06,sbk+06,bgc+06}, the
total energy of \grb\ is consistent with predictions for both
scenarios.

The second major consequence of beaming is that the true event rate is
{\it higher} than the observed rate by the inverse of the beaming
factor (i.e., $R_{\rm true}=f_{b}^{-1}R_{\rm obs}$). Thus, beaming
provides essential information for understanding the relation to
various progenitor systems and is of particular interest since the
NS-NS/NS-BH merger rates, which are a critical input for estimates of
Advanced LIGO gravitational wave detections, are highly uncertain
(e.g., \citealt{aaa+10,mb12}). The current estimated {\it observed}
short GRB volumetric rate is $\sim\!10$ Gpc$^{-3}$ yr$^{-1}$
\citep{ngf06}. The estimated NS-NS merger rate is much higher:
$\sim\!200$-$3000$ Gpc$^{-3}$ yr$^{-1}$ \citep{kkl+04,ngf06}.

The discrepancy in these rates can be explained if short GRBs have
typical $\theta_j \sim 8^{\circ}$ ($f_b^{-1} \sim 100$; see also
\citealt{mb12}). The determination of \grb's opening angle of
$3-8^{\circ}$ ($f_b^{-1} = 100-730$), along with the small but
increasing sample of opening angle constraints for short GRBs, implies
that at least a fraction of these events are significantly beamed and
that the true rate of short GRBs is at least $\sim 100-1000$
Gpc$^{-3}$ yr$^{-1}$. With a few additional opening angle
measurements, this value can be significantly improved. Other proposed
progenitor models, e.g., WD-WD mergers or accretion-induced collapse
of a WD/NS \citep{qwc+98,lwc+06,mqt08} have estimated rates of
$\lesssim 1000$ Gpc$^{-3}$ yr$^{-1}$ and $\sim 0.1-100$ Gpc$^{-3}$
yr$^{-1}$, respectively \citep{lr07,dmq+10}, so if a large fraction of
short GRBs have opening angles of $\lesssim 25^{\circ}$, these systems
may not contribute significantly to the progenitor population.

\section{Conclusions and Future Work}

We have presented observations of \grb, utilizing extensive coverage
in the X-rays with \swift/XRT, XMM and \chandra\ to uncover a temporal
break, most naturally explained as a jet break. Our limit on the radio
afterglow from EVLA combined with the inference that $\nu_c<\nu_X$
leads to a robust range on the circumburst density of $\sim 0.01-0.1$
cm$^{-3}$. The jet break time of $\approx 2$ days leads to an opening
angle of $3-8^{\circ}$, depending on the redshift and equipartition
fractions, which leads to beaming-corrected energies of $E_{\gamma}
\approx (2-3) \times 10^{48}$ erg, $E_K \approx (0.3-2) \times
10^{49}$ erg and $E_{\rm tot} \approx (0.5-2) \times 10^{49}$
erg. This result, along with the previous jet break constraints for
GRBs\,051221A and 050724A suggests that there may be a spread in true
energy release, $\sim 10^{48}-10^{50}$ erg for short GRBs
\citep{bpc+05,gbp+06,sbk+06,bgc+06}.

Furthermore, our optical observations provide a limit on the afterglow
and enabled the discovery of a putative host galaxy with $i \approx
24.3$ mag. A comparison of the X-ray and optical data at $\delta t =
17.7$ hours provides a lower limit on the host galaxy extinction of
$A_V^{\rm host} \gtrsim 0.2-0.6$ mag. This is consistent with the high
intrinsic column density from X-ray absorption when compared to the
mean for the short GRB population.

\grb\ demonstrates that rapid multi-wavelength follow-up is vital to
our understanding of the basic properties of short GRBs: the geometry,
energetics, and circumburst densities. In particular, the search for
jet breaks on timescales of $\gtrsim$ few days is imperative for
placing meaningful constraints on the opening angle
distribution. Ideally, the detection of breaks in both optical and
X-ray data leads to an unambiguous and tight constraint on the opening
angle; however, optical afterglows are only detected in $\sim\!30\%$
of \swift\ short GRBs, while X-ray afterglows have been detected in
$\sim\!70\%$. Furthermore, optical afterglows are intrinsically faint
and subject to host galaxy contamination, making long-term monitoring
highly challenging. Therefore, the jet break search is optimized in
the X-ray band where the burst is not subject to such contamination
and the afterglow brightness is virtually independent of the typically
low circumburst densities. The X-rays also allow for a measurement of
the kinetic energy of the outflow. Deep radio limits provide
additional constraints on the circumburst density and energy. The EVLA
upgrade is now enabling us to probe events with relatively low energy
scales of $\sim 10^{48}$ erg and densities of $\lesssim 10^{-2}$
cm$^{-2}$.

The collimation of short GRBs will undoubtedly further our knowledge
of their true energetics and rates. While the former provides
information on the explosion and energy extraction mechanisms, the
latter is crucial for understanding the relation to various progenitor
systems (e.g., NS-NS mergers). Significant improvement on the
estimated short GRB observed rate of $\sim 10$ Gpc$^{-3}$ yr$^{-1}$
\citep{ngf06} will have a critical impact on estimates for coincident
short GRB-gravitational wave detections in the era of Advanced
LIGO/VIRGO \citep{aaa+10}. The uncertainty in the observed short GRB
rate is dominated by the uncertainty in the beaming fraction and with
only a handful of short GRB opening angles measured to date, the
discovery of even a few additional jet breaks in the coming years
will enable significant progress.

\acknowledgments

We thank D.~Finkbeiner for helpful discussions. The Berger GRB group
at Harvard is supported by the National Science Foundation under Grant
AST-1107973.  Partial support was also provided by the National
Aeronautics and Space Administration through Chandra Award Number
GO1-12072X issued by the Chandra X-ray Observatory Center, which is
operated by the Smithsonian Astrophysical Observatory for and on
behalf of the National Aeronautics Space Administration under contract
NAS8-03060.  Additional support was provided by NASA/Swift AO6 grant
NNX10AI24G. Observations were obtained with the EVLA (program 10C-145)
operated by the National Radio Astronomy Observatory, a facility of
the National Science Foundation operated under cooperative agreement
by Associated Universities, Inc.  This paper includes data gathered
with the 6.5 meter Magellan Telescopes located at Las Campanas
Observatory, Chile.  This work is based in part on observations
obtained at the Gemini Observatory, which is operated by the
Association of Universities for Research in Astronomy, Inc., under a
cooperative agreement with the NSF on behalf of the Gemini
partnership: the National Science Foundation (United States), the
Science and Technology Facilities Council (United Kingdom), the
National Research Council (Canada), CONICYT (Chile), the Australian
Research Council (Australia), Ministério da Ciência, Tecnologia e
Inovação (Brazil) and Ministerio de Ciencia, Tecnología e Innovación
Productiva (Argentina).

\clearpage
\begin{deluxetable}{lccccc}
\tabletypesize{\small}
\tablecolumns{6}
\tabcolsep0.05in\small
\tablewidth{0pc}
\tablecaption{\grb\ X-ray Spectral Fit Parameters
\label{tab:xrayspec}}
\tablehead {
\colhead {Telescope} &
\colhead {Detector}            &
\colhead {$\delta t$}           &
\colhead {N$_{\rm H,int}^{ab}$}          &
\colhead {$\Gamma^{ab}$}         &
\colhead {C-stat$_\nu$/d.o.f.} \\
\colhead {}          &
\colhead {}              &
\colhead {(ks)}       &
\colhead {($10^{22}$ cm$^{-2}$)}       &
\colhead {}   &  
\colhead {}     
}
\startdata
\swift        & XRT    &  $0.08-60$   & $1.0 \pm 0.3$          & $2.2 \pm 0.5$ & $0.86/188$ \\
XMM           & EPIC-PN & $61.4-76.8$ & $0.65^{+0.21}_{-0.23}$ & $2.0 \pm 0.4$ & $1.0/256$ \\
{\it Chandra} & ACIS-S & $250.5-268.5$ & $0.4^{+2.3}_{-0.4}$    & $1.1^{+2.7}_{-0.8}$ & $0.32/661$  \\
\swift+XMM    & XRT+EPIC-PN & see above  & $0.75^{0.20}_{-0.18}$  & $2.0 \pm 0.3$ & $0.94/446$
\enddata
\tablecomments{$^a$ These values assume a Galactic column density of
$N_{{\rm H,gal}}=6.9 \times 10^{20}$ cm$^{-2}$ \citep{kbh+05}, using an {\tt XSPEC} model of $tbabs \times ztbabs \times
pow$ at $z=0$. \\
$^b$ Uncertainties correspond to a $90\%$ confidence level.
}
\end{deluxetable}

\clearpage
\begin{deluxetable}{lcc}
\tabletypesize{\scriptsize}
\tablecolumns{3}
\tabcolsep0.05in\footnotesize
\tablewidth{0pc}
\tablecaption{GRB\,111020A X-ray Observations
\label{tab:xrayobs}}
\tablehead {
\colhead {$\delta t$}            &
\colhead {Time Bin Duration}       &
\colhead {Unabs. Flux ($0.3-10$ keV)} \\
\colhead {(s)}          &
\colhead {(s)}              &
\colhead {(erg cm$^{-2}$ s$^{-1}$)}        
}
\startdata
\multicolumn{3}{c}{{\it Swift/XRT}}  \\
\noalign{\smallskip}
$6.18 \times 10^{1^a}$ & $7.44 \times 10^{0}$ & $(2.38 \pm 0.79) \times 10^{-10}$ \\
$1.35 \times 10^{2}$ & $4.64 \times 10^{1}$ & $(3.80 \pm 1.03) \times 10^{-11}$  \\
$2.66 \times 10^{2}$ & $1.71 \times 10^{2}$ & $(1.64 \pm 0.42) \times 10^{-11}$  \\
$4.15 \times 10^{2}$ & $1.26 \times 10^{2}$ & $(2.43 \pm 0.64) \times 10^{-11}$  \\
$5.96 \times 10^{2}$ & $2.36 \times 10^{2}$ & $(1.03 \pm 0.26) \times 10^{-11}$  \\
$7.97 \times 10^{2}$ & $1.66 \times 10^{2}$ & $(1.83 \pm 0.49) \times 10^{-11}$  \\
$1.14 \times 10^{3}$ & $5.20 \times 10^{2}$ & $(8.85 \pm 1.72) \times 10^{-12}$  \\
$5.94 \times 10^{3}$ & $2.46 \times 10^{3}$ & $(1.65 \pm 0.33) \times 10^{-12}$  \\
$1.17 \times 10^{4}$ & $2.46 \times 10^{3}$ & $(1.07 \pm 0.27) \times 10^{-12}$  \\
$1.94 \times 10^{4}$ & $6.38 \times 10^{3}$ & $(9.19 \pm 2.28) \times 10^{-13}$  \\
$2.58 \times 10^{4}$ & $6.28 \times 10^{3}$ & $(1.19 \pm 0.32) \times 10^{-12}$  \\
$3.19 \times 10^{4}$ & $5.90 \times 10^{3}$ & $(1.05 \pm 0.27) \times 10^{-12}$  \\
$4.29 \times 10^{4}$ & $1.61 \times 10^{4}$ & $(8.36 \pm 2.41) \times 10^{-13}$  \\
$1.26 \times 10^{5}$ & $1.51 \times 10^{5}$ & $(1.63 \pm 0.55) \times 10^{-13}$  \\
$3.09 \times 10^{5}$ & $2.14 \times 10^{5}$ & $(1.11 \pm 0.42) \times 10^{-13}$  \\
\noalign{\smallskip}
\hline
\noalign{\smallskip}
\multicolumn{3}{c}{{\it XMM/EPIC-PN}} \\
\noalign{\smallskip}
$6.91 \times 10^{4}$ & $1.35 \times 10^{4}$ & $(2.66 \pm 0.19) \times 10^{-13}$  \\
\noalign{\smallskip}
\hline
\noalign{\smallskip}
\multicolumn{3}{c}{{\it Chandra/ACIS-S}} \\
\noalign{\smallskip}
$2.61 \times 10^{5}$ & $1.98 \times 10^{4}$ & $(5.96 \pm 0.89) \times 10^{-14}$  \\
$8.84 \times 10^{5}$ & $1.98 \times 10^{4}$ & $<8.95 \times 10^{-15}$ 
\enddata
\tablecomments{Upper limits are $3\sigma$. \\
$^a$ These points were excluded from the broken power law fit. \\
}
\end{deluxetable}

\clearpage
\begin{deluxetable}{lcccccccccccc}
\tabletypesize{\scriptsize}
\tablecolumns{13}
\tabcolsep0.0in\scriptsize
\tablewidth{0pc}
\tablecaption{GRB\,111020A Optical Photometry
\label{tab:opt_phot}}
\tablehead {
\colhead {Date} &
\colhead {$\Delta t$}            &
\colhead {Telescope}           &
\colhead {Instrument}          &
\colhead {Filter}          &
\colhead {Exposures}       &
\colhead {$\theta_{\rm FWHM}$} &
\colhead {Afterglow$^{ab}$}           &
\colhead {$F_\nu$$^{ab}$}                 &
\colhead {G1$^a$}      &
\colhead {G2$^a$}  &
\colhead {G3$^a$}  &
\colhead {$A_{\lambda}$}    \\
\colhead {(UT)}          &
\colhead {(d)}          &
\colhead {}              &
\colhead {}       &
\colhead {}       &
\colhead {(s)}              &
\colhead {(arcsec)}      &
\colhead {(AB mag)}           & 
\colhead {($\mu$Jy)}     &
\colhead {(AB mag)} &
\colhead {(AB mag)}   &
\colhead {(AB mag)}  &
\colhead {(AB mag)}  
}
\startdata
2011 October 21.01 & $0.74$ & Magellan/Clay & LDSS3 & $r$ & $3 \times 360$ & $0.62$  & $>23.4$ & $<1.56$ & $>23.4$     & $21.12 \pm 0.09$  & $>23.4$           & $0.987$  \\
2011 October 21.01 & $0.74$ & Gemini-S & GMOS  & $i$ & $9 \times 180$  & $0.74$ & $>24.4$ & $<0.63$ &  &  &  & $0.734$ \\
2011 October 22.01 & $1.74$ & Gemini-S & GMOS  & $i$ & $11 \times 180$ & $0.67$ &         &        &  &   &   & $0.734$ \\
2011 October 21.01+22.01  &        & Gemini-S & GMOS  & $i$ & $20 \times 180$ & $0.72$ &         &        & $23.89 \pm 0.17$ & $21.91 \pm 0.05$  & $24.27 \pm 0.16$  & $0.734$
\enddata
\tablecomments{$^a$ These values have been corrected for Galactic extinction $A_\lambda$ \citep{sf11}. \\
$^b$ Limits are $3\sigma$.
}
\end{deluxetable}

\clearpage
\begin{deluxetable}{lcc}
\tabletypesize{\small}
\tablecolumns{3}
\tabcolsep0.05in\small
\tablewidth{0pc}
\tablecaption{Physical Parameters of \grb\
\label{tab:param}}
\tablehead {
\colhead {Parameter } &
\colhead {Case I [$z=0.5, \epsilon_e=\epsilon_B=1/3$]}   &
\colhead {Case II [$z=1.5, \epsilon_e=\epsilon_B=0.1$]}                   
}
\startdata
$t_j$        & $2.0 \pm 0.5$ days$^a$ & $2.0 \pm 0.5$ days$^a$ \\
$E_{\gamma, {\rm iso}}$ & $2.1 \times 10^{50}$ erg & $1.9 \times 10^{51}$ erg \\
$E_{\rm K,iso}$  & $3.7 \times 10^{50}$ erg & $1.2 \times 10^{52}$ erg \\
$n_0$          & $0.01-0.06$ cm$^{-3}$ & $0.008$ cm$^{-3}$ \\
$\theta_j$   & $7-8^{\circ}$ & $3^{\circ}$   \\
$f_b$        & $0.007-0.01$ & $0.001$ \\
$E_{\gamma}$ & $2 \times 10^{48}$ erg & $3 \times 10^{48}$ erg \\
$E_{K}$  & $(3-4) \times 10^{48}$ erg & $2 \times 10^{49}$ erg \\
$E_{\rm tot}$ & $(5-6) \times 10^{48}$ erg & $2 \times 10^{49}$ erg \\
$\eta_\gamma$ & $0.3-0.4$ & $0.15$ 
\enddata
\tablecomments{$^a$ Uncertainties correspond to a $1\sigma$ confidence level.
}
\end{deluxetable}

\clearpage
\begin{deluxetable}{lccc}
\tabletypesize{\small}
\tablecolumns{4}
\tabcolsep0.05in\small
\tablewidth{0pc}
\tablecaption{Intrinsic X-ray Column Density of Hydrogen, $N_{\rm H,int}$ for \swift\ Short GRBs
\label{tab:NH}}
\tablehead {
\colhead {GRB} &
\colhead {$z$}   &
\colhead {$N_{\rm H,int}$}      &
\colhead {$\sigma$ above zero} \\
\colhead {} &
\colhead {} &
\colhead {($10^{21}$ cm$^{-2}$)} &
\colhead {}
}
\startdata
050724  & $0.258$ & $3.20^{+0.97}_{-0.86}$ & $5.7$ \\
051210  & \nod    & $<0.54$                &       \\
051221A & $0.547$ & $1.92^{+0.73}_{-0.68}$ & $4.5$ \\
060313  & \nod    & $0.45^{+0.36}_{-0.33}$ & $2.1$ \\
060801  & $1.131$ & $3.02^{+2.22}_{-1.88}$ & $2.4$ \\
061006  & $0.438$ & $<2.04$                &       \\
061201  & \nod    & $0.94^{+0.60}_{-0.53}$ & $2.7$ \\
070714B & $0.923$ & $3.89^{+1.87}_{-1.61}$ & $4.2$ \\
070724A & $0.457$ & $<1.89$                &       \\
071227  & $0.383$ & $2.84^{+0.72}_{-0.65}$ & $6.8$ \\
080123  & \nod    & $1.12^{+0.28}_{-0.26}$ & $6.8$ \\
080905A & $0.122$ & $2.04^{+1.58}_{-1.33}$ & $2.3$ \\
090510  & $0.903$ & $<0.80$                &       \\
090515  & \nod    & $0.56^{+0.30}_{-0.27}$ & $3.2$ \\
090607  & \nod    & $<0.79$                &       \\
091109B & \nod    & $<1.58$                &       \\
100117A & $0.915$ & $4.10^{+3.41}_{-2.71}$ & $2.2$ \\
100702A & \nod    & $4.37^{+3.67}_{-3.05}$ & $2.1$ \\
101219A & $0.718$ & $6.61^{+3.73}_{-2.82}$ & $3.3$ \\
110112A & \nod    & $<0.92$                &       \\
111020A & \nod    & $7.50^{+2.0}_{-1.8}$   & $6.5$ \\
111117A & \nod    & $1.84^{+1.28}_{-1.05}$ & $2.6$ \\
111121A & \nod    & $2.41^{+0.82}_{-0.74}$ & $5.1$ 
\enddata
\tablecomments{Errors and upper limits quoted correspond to a $90\%$ confidence level;
$z=0$ is assumed when the redshift is not known.
}
\end{deluxetable}

\clearpage

\clearpage

\clearpage

\clearpage

\clearpage

\clearpage

\end{document}